\documentclass[12pt,preprint]{aastex}
\pdfoutput=1

\usepackage{amsmath}
\usepackage{amsfonts}
\usepackage{dsfont}
\usepackage{amsxtra}
\usepackage{hyperref}
\usepackage{amssymb}

\usepackage{multirow}
\usepackage{url}

\usepackage{graphicx,epsfig}
\usepackage{dcolumn}
\usepackage{bm}

\usepackage{color}
\usepackage{longtable}
\usepackage{threeparttable}
\newcommand{\Fermi}{\textit{Fermi}}

\newcommand{\onepic}{0.4}
\newcommand{\twopic}{0.35}

\begin{document}

\title{Search for Early Gamma-ray Production in Supernovae Located in a Dense Circumstellar Medium with the \Fermi\,LAT}

\author{
M.~Ackermann\altaffilmark{1}, 
I.~Arcavi\altaffilmark{2,3},
L.~Baldini\altaffilmark{4}, 
J.~Ballet\altaffilmark{5}, 
G.~Barbiellini\altaffilmark{6,7}, 
D.~Bastieri\altaffilmark{8,9}, 
R.~Bellazzini\altaffilmark{10}, 
E.~Bissaldi\altaffilmark{11}, 
R.~D.~Blandford\altaffilmark{12}, 
R.~Bonino\altaffilmark{13,14}
% ...
}
\altaffiltext{1}{Deutsches Elektronen Synchrotron DESY, D-15738 Zeuthen, Germany}
\altaffiltext{2}{Las Cumbres Observatory Global Telescope Network, 6740 Cortona Dr., Suite 102, Goleta, CA 93117, USA}
\altaffiltext{3}{Kavli Institute for Theoretical Physics, University of California, Santa Barbara, CA 93106, USA}
\altaffiltext{4}{Universit\`a di Pisa and Istituto Nazionale di Fisica Nucleare, Sezione di Pisa I-56127 Pisa, Italy}
\altaffiltext{5}{Laboratoire AIM, CEA-IRFU/CNRS/Universit\'e Paris Diderot, Service d'Astrophysique, CEA Saclay, 91191 Gif sur Yvette, France}
\altaffiltext{6}{Istituto Nazionale di Fisica Nucleare, Sezione di Trieste, I-34127 Trieste, Italy}
\altaffiltext{7}{Dipartimento di Fisica, Universit\`a di Trieste, I-34127 Trieste, Italy}
\altaffiltext{8}{Istituto Nazionale di Fisica Nucleare, Sezione di Padova, I-35131 Padova, Italy}
\altaffiltext{9}{Dipartimento di Fisica e Astronomia ``G. Galilei'', Universit\`a di Padova, I-35131 Padova, Italy}
\altaffiltext{10}{Istituto Nazionale di Fisica Nucleare, Sezione di Pisa, I-56127 Pisa, Italy}
\altaffiltext{11}{Istituto Nazionale di Fisica Nucleare, Sezione di Bari, 70126 Bari, Italy}
\altaffiltext{12}{W. W. Hansen Experimental Physics Laboratory, Kavli Institute for Particle Astrophysics and Cosmology, Department of Physics and SLAC National Accelerator Laboratory, Stanford University, Stanford, CA 94305, USA}
\altaffiltext{13}{Istituto Nazionale di Fisica Nucleare, Sezione di Torino, I-10125 Torino, Italy}
\altaffiltext{14}{Dipartimento di Fisica Generale ``Amadeo Avogadro" , Universit\`a degli Studi di Torino, I-10125 Torino, Italy}
% ...

\begin{abstract}
Supernovae (SNe) exploding in a dense circumstellar medium (CSM) are hypothesized to accelerate cosmic rays in collisionless shocks and emit GeV $\gamma$ rays and TeV neutrinos on a time scale of several months. We perform the first systematic search for $\gamma$-ray emission in \Fermi\,LAT data in the energy range from $100\,$MeV to $300\,$GeV from the ensemble of 147 SNe Type IIn exploding in dense CSM. We search for a $\gamma$-ray excess at each SNe location in a one year time window. 
In order to enhance a possible weak signal, we simultaneously study the closest and optically brightest sources of our sample in a joint-likelihood analysis in three different time windows (1\,year, 6\,months and 3\,months). For the most promising source of the sample, SN\,2010jl (PTF10aaxf), we repeat the analysis with an extended time window lasting 4.5 years. We do not find a significant excess in $\gamma$ rays for any individual source nor for the combined sources and provide model-independent flux upper limits for both cases. In addition, we derive limits on the $\gamma$-ray luminosity and the ratio of $\gamma$-ray-to-optical luminosity ratio 
as a function of the index of the proton injection spectrum assuming a generic $\gamma$-ray production model. 
Furthermore, we present detailed flux predictions based on multi-wavelength observations and the corresponding flux upper limit at  $95\%$ confidence level (CL) for the source SN\,2010jl (PTF10aaxf).
\end{abstract}



\section{Introduction}
\label{sec:Intro}

The Large Area Telescope (LAT) on-board the \Fermi\,Gamma-ray Space Telescope mission unanticipatedly detected $\gamma$-ray emission from five Galactic novae~\citep{Abdo:2010he,Hill:2013fia,atel1,atel2}. The origin of the $\gamma$-ray emission is still unclear. Shocks produced by expansion of the nova shell into the wind provided by the companion star or internal shocks within the ejecta might be responsible for acceleration of particles to relativistic energies and ensuing high-energy $\gamma$-ray emission. A similar mechanism but with much larger energy output is hypothesized to produce $\gamma$ rays in supernovae (SNe) yielding potentially detectable $\gamma$-ray emission even from extragalactic sources. \citet{Murase:2010cu,Murase:2013kda} and~\citet{Katz:2011zx} showed that if the SN progenitor is surrounded by an optically thick circumstellar medium (CSM), then a collisionless shock is necessarily formed after the shock breakout. The collisionless shock may accelerate protons and electrons to high energies, which emit photons from the radio-submillimeter through GeV energies and TeV neutrinos.
Such conditions appear in shocks propagating through dense circumstellar matter (e.g., wind).
Recently several candidates for such SNe powered by interactions with a dense CSM were found \citep[e.g.,][]{2007ApJ...659L..13O,2014ApJ...781...42O,2009AJ....137.3558S,Zhang2012}
and some superluminous supernovae were suggested to be powered by interactions \citep[e.g.,][]{Quimby:2009ps,Chevalier:2011ha}. Such interaction-powered supernovae may also be Pevatrons, implying their importance for the origin of the knee structure in the cosmic-ray spectrum~\citep{Murase:2013kda,2003A&A...409..799S}. Both $\gamma$ rays and neutrinos originate from pp and p$\gamma$ interactions producing pions, which in the neutral case decay to $\gamma$ rays and in the charged case produce neutrinos in the decay chain. Thus, the initial neutrino and $\gamma$-ray spectra have the same shape. Contrary to neutrinos, $\gamma$ rays might be affected by absorption in the CSM and/or two-photon annihilation with low-energy photons produced at the forward shock~\citep{Murase:2010cu}. However, arguments made in~\citet{Murase:2013kda} suggest that GeV $\gamma$ rays can escape the system without severe attenuation, if the shock velocity is in the right range, especially late after the shock breakout. 

Motivated by the fact that the LAT has detected $\gamma$-ray emission from novae, we are presenting the first systematic search for $\gamma$-ray emission from Type IIn SNe in \Fermi\,LAT data from $100\,$MeV to $300\,$GeV. Considering current theoretical uncertainties we are aiming for a model independent search. SNe positions and explosion times are given by optical surveys such as the Palomar Transient Factory~\citep[PTF --][]{Law:2009ys,Rau2009}. 

We present the sample of SNe used in the $\gamma$-ray data analysis in Section~\ref{sec:SNCat}. Section~\ref{sec:Fermi} describes the \Fermi\,LAT data analysis followed by an interpretation of our results in Section~\ref{sec:Interpretation} and conclusions in Section~\ref{sec:Conclusions}.

\section{SNe Sample}
\label{sec:SNCat}

\begin{figure}[htbp]
\begin{center}
\includegraphics[scale=\onepic]{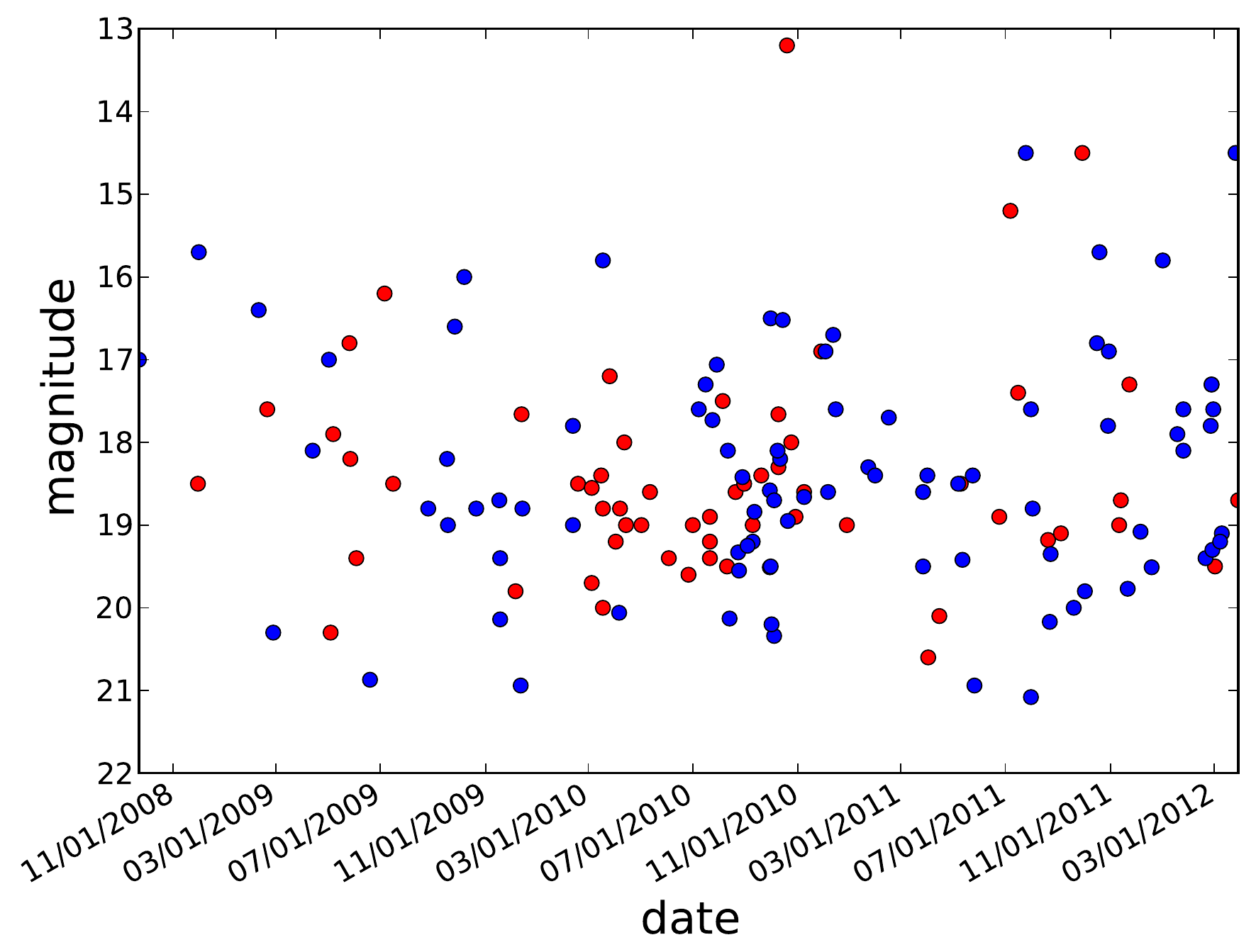}
\noindent
\caption{\small 
Apparent $R$-band peak (detection) magnitude as a function of the peak (detection) time shown in red (blue) for all 147 SNe in our sample. For some SNe the peak time and magnitude is not determined; in those cases we use the detection time and magnitude.}
\label{fig:SNMag}
\end{center}
\vspace{1mm}
\end{figure}

Type IIn and Type Ibn SNe are the best candidates to be interacting with a dense CSM. 
Their long-lasting bright optical light curves are believed to be powered by the interaction of the ejecta with massive CSM~\citep{2012ApJ...759..108S}. SNe of these types are often accompanied by precursor mass-ejection events~\citep{Ofek:2014ifa}. 
Here we mainly use the PTF SN sample along with publicly available Type IIn
SNe discovered since the launch of \Fermi\,in 2008. 
Appendix~\ref{sec:AppendixA} lists all $147$ SNe of this sample that we consider in our $\gamma$-ray search, i.e., all sources with an estimated explosion time later than 2008 August 4 and before 2012 May 1 (this is one year before the end of the studied $\gamma$-ray data sample). The apparent $R$-band peak magnitude ($m$) as a function of the peak time is shown in Fig.~\ref{fig:SNMag}. Note that throughout this paper we refer to $m$ as the peak magnitude; for sources where the peak magnitude is not determined we use the discovery magnitude instead. 
The subsample of bright ($m<16.5$) and/or nearby (with a redshift $z<0.015$) SNe used for the joint likelihood analysis is detailed in Table~\ref{tab:closeSNshort}.

\begin{center}
{\footnotesize\begin{threeparttable}
\begin{tabular}{ l  c c c c c c }
\hline
\hline
Name & RA ($^\circ$)$^{\star}$ & Dec ($^\circ$)$^{\star}$ & Date  & z & m & TS (p-value) \\
\hline
\hline
SN2008gm & 348.55 & -2.78 & 2008-10-22\tnote{$\dagger$}  & 0.012 & 17.00\tnote{$\ddagger$}  & 3.2 (0.169) \\
\hline
SN2008ip & 194.46 & 36.38 & 2008-12-31\tnote{$\dagger$}  & 0.015 & 15.70\tnote{$\ddagger$}  & 0.0 (0.572) \\
\hline
SN2009au & 194.94 & -29.60 & 2009-03-11\tnote{$\dagger$}  & 0.009 & 16.40\tnote{$\ddagger$}  & 0.0 (0.572) \\
\hline
PTF10ujc & 353.63 & 22.35 & 2009-08-05 & 0.032 & 16.20 & 0.0 (0.572) \\
\hline
SN2009kr & 78.01 & -15.70 & 2009-11-06\tnote{$\dagger$}  & 0.006 & 16.00\tnote{$\ddagger$}  & 4.7 (0.104) \\
\hline
SN2010bt & 192.08 & -34.95 & 2010-04-17\tnote{$\dagger$}  & 0.016 & 15.80\tnote{$\ddagger$}  & 14.4 (0.0065) \\
\hline
PTF10aaxf & \multirow{2}{*}{145.72} & \multirow{2}{*}{9.50} & \multirow{2}{*}{2010-11-18} & \multirow{2}{*}{0.011} & \multirow{2}{*}{13.20} & \multirow{2}{*}{7.1 (0.039)} \\
SN2010jl & & & & & & \\
\hline
PTF10aaxi & \multirow{2}{*}{94.13} & \multirow{2}{*}{-21.41} & \multirow{2}{*}{2010-11-23} & \multirow{2}{*}{0.010} & \multirow{2}{*}{18.00} & \multirow{2}{*}{0.0 (0.572)} \\
SN2010jp & & & & & & \\
\hline
SN2011A & 195.25 & -14.53 & 2011-01-02\tnote{$\dagger$}  & 0.009 & 16.90\tnote{$\ddagger$}  & 0.0 (0.572) \\
\hline
PTF11iqb & 8.52 & -9.70 & 2011-08-06 & 0.013 & 15.20 & 0.3 (0.469) \\
\hline
SN2011fh & 194.06 & -29.50 & 2011-08-24\tnote{$\dagger$}  & 0.008 & 14.50\tnote{$\ddagger$}  & 1.9 (0.262) \\
\hline
PSNJ10081059+5150570 & \multirow{2}{*}{152.04} & \multirow{2}{*}{51.85} & \multirow{2}{*}{2011-10-29} & \multirow{2}{*}{0.004} & \multirow{2}{*}{14.50} & \multirow{2}{*}{0.0 (0.572)} \\
SN2011ht & & & & & & \\
\hline
PTF11qnf & 86.23 & 69.15 & 2011-11-01\tnote{$\dagger$}  & 0.014 & 19.80\tnote{$\ddagger$}  & 1.4 (0.320) \\
\hline
SN2011hw & 336.56 & 34.22 & 2011-11-18\tnote{$\dagger$}  & 0.023 & 15.70\tnote{$\ddagger$}  & 0.0 (0.572) \\
\hline
SN2012ab & 185.70 & 5.61 & 2012-01-31\tnote{$\dagger$}  & 0.018 & 15.80\tnote{$\ddagger$}  & 0.0 (0.572) \\
\hline
PSNJ18410706-4147374 & \multirow{2}{*}{280.28} & \multirow{2}{*}{-41.79} & \multirow{2}{*}{2012-04-25}\tnote{$\dagger$}  & \multirow{2}{*}{0.019} & \multirow{2}{*}{14.50}\tnote{$\ddagger$}  & \multirow{2}{*}{0.0 (0.572)} \\
SN2012ca & & & & & & \\
\hline
\hline
\end{tabular}
\begin{tablenotes}
\item[$\star$] Epoch J2000.0
\item[$\dagger$] Discovery date
\item[$\ddagger$] Discovery magnitude
\end{tablenotes}
\caption{List of nearby and/or bright SNe -- with redshift $z<0.015$ and/or R-band magnitude $m<16.5$. The colums contain the name of the SN, its direction in equatorial coordinated (right ascension, RA, and declination, Dec), its peak date and peak R-band magnitude, its redshift, its test statistic (TS) and p-value. See Section~\ref{sourceBySource} for details on the TS and p-value calculation. Note that if the peak date and magnitude are not available in the catalog, the discovery date and magnitude are quoted instead.}
\label{tab:closeSNshort}
\end{threeparttable}
}\end{center}

\section{ \textit{Fermi} LAT $\gamma$-ray Data Analysis}
\label{sec:Fermi}

The \Fermi\,LAT is a pair-conversion telescope, sensitive to $\gamma$ rays with energies from $20\,$MeV to greater than $300\,$GeV~\citep{2009ApJ...697.1071A}. It has a large field of view and has been scanning the entire sky every few hours for the last 6 years. Thus it is very well suited 
for searches for transient $\gamma$-ray signals on the timescale of months. Bright SN events may be detectable at distances $d<30$\,Mpc \citep{Murase:2010cu} depending on the properties of the source. \citet{Margutti:2013pfa} searched for $\gamma$ rays from a single SN in the case of SN2009ip located at a distance of $24\,$Mpc. No $\gamma$-ray excess  was identified in \Fermi\,LAT data at the SN position; this is  consistent with the picture of ejecta colliding with a compact and dense but low-mass shell of material. For a detection of a single source, closer and/or brighter SNe are needed (i.e., reached by larger dissipation and larger CSM masses). The properties of the ejecta and CSM can be estimated from multi-wavelength observations in a few cases \citep[e.g., SN2009ip,][]{2013ApJ...768...47O}, but are uncertain or not known in most cases. 

In this analysis we use 57 months of \Fermi\,LAT data recorded between 2008 August 4 and 2013 May 1 (\Fermi\,Mission Elapsed Time 239557418--389092331 s),
restricted to the Pass 7 Reprocessed Source class\footnote{\url{http://fermi.gsfc.nasa.gov/ssc/data/analysis/documentation/Pass7REP_usage.
html}}. We select the standard good time intervals (e.g., excluding time intervals when the field of view of the LAT intersected the Earth). The Pass 7 Reprocessed data benefit from an updated calibration that improves the energy measurement and event-direction reconstruction accuracy at energies above 1 GeV~\citep{2013arXiv1304.5456B}. To minimize the contamination from the 
$\gamma$ rays produced in the upper atmosphere, we select events with zenith angles $<100^\circ$. We perform a binned analysis (i.e., binned in space and energy) using the standard \Fermi\,LAT ScienceTools package version v09r32p05 available from the \Fermi\,Science Support Center\footnote{\url{http://fermi.gsfc.nasa.gov/ssc/data/analysis/}} (FSSC) using the P7REP\_SOURCE\_V15 instrument response functions. We analyze data in the energy range of 100\,MeV to 300\,GeV binned into 20 logarithmic energy intervals. For each source we select a $20^\circ \times 20^\circ$ region of interest (ROI) centered on the source localization binned in $0\fdg2$ size pixels. The binning is applied in celestial coordinates and an Aitoff projection was used.
 
We use four different approaches in our analysis:
\begin{enumerate}
\item 
We perform a likelihood analysis to search for $\gamma$-ray excesses that are consistent with originating from a point source coincident with the position of each Type IIn SNe in our sample over a 1-year time scale. We assume that their $\gamma$-ray emission follows a power-law spectrum. This approach is sensitive to single bright sources.
\item 
In a model-independent approach (i.e., no prior assumption on the SN $\gamma$-ray spectral shape) we compute the likelihood in bins of energy (bin-by-bin likelihood). We use the bin-by-bin likelihood to evaluate $95$\% CL flux upper limits in 20 energy bins for the 16 closest and optically brightest SNe in our sample.
\item In order to increase the sensitivity for a weak signal, we combine individual sources in a joint likelihood analysis using the composite likelihood tool, \textit{Composite2}, of the \Fermi\, Science Tools.
\item 
We repeat the joint likelihood analysis using the composite likelihood tool, but limit the sample to those Type IIn SNe that exhibit additional indications of strong interactions with their CSM. Not all Type IIn SNe might be surrounded by a massive CSM. This clean sample of SNe with confirmed massive CSM might produce a strong $\gamma$-ray signal 
and should provide an enhanced signal-to-background ratio.
\end{enumerate}
Accurate SN positions are given by optical localizations. Theoretical predictions of the duration of the $\gamma$-ray emission are uncertain and motivate a search in several time windows. We test three different time windows: $\Delta T = 1$\,year, 6\,months and 3\,months. 
The optical light curve is produced by the interaction of the SN ejecta with the dense CSM and is thus correlated with the expected $\gamma$-ray emission. 
Most of the $\gamma$-ray emission is expected during the interactions after the shock breakout. The optical light curve peak is reached around the end of the breakout~\cite[see e.g.,][]{Ofek:2010kq}. 
We collected the SN properties from the PTF sample, Astronomer's Telegrams (ATels)\footnote{\url{http://www.astronomerstelegram.org/}} and Central Bureau for Astronomical Telegrams (CBETs)\footnote{\url{http://www.cbat.eps.harvard.edu/cbet/RecentCBETs.html}}. Most PTF sources are unpublished and the other events were drawn from ATEL and CBET. Full details and final analysis of the PTF SN IIn sample will be provided in a forthcoming publication. In some cases the known SN properties include the optical flux peak time while in other cases this information is missing and only the optical detection time is available. To account for the uncertainty in the determination of the peak time and to make sure no early $\gamma$-ray emission is missed, we start the time window $30$\,days before the peak time (or the detection time in case the peak time is not provided). In the case of the three novae, the reported $\gamma$-ray light curves~\citep[see Fig. 1 in][]{Hill:2013fia} have very similar durations justifying a similar time window for all sources. However, the duration of the novae detected by \Fermi\,were $\sim20$\,days, while SN\,IIn typically last longer, $\mathcal{O}$(100\,days - 1\,year).

\subsection{Source Specific Analysis}
\label{sourceBySource}
We analyze the $20^\circ \times 20^\circ$ ROI around each source in our SN sample in a 1-year time window in a binned likelihood analysis. We construct a model whose free parameters are fitted to the data in the ROI. This model includes a point-like source at the SN position; its $\gamma$-ray spectrum is represented as a power-law function with both index and normalization free to vary. In addition we have to model the point sources in the ROI and the diffuse $\gamma$-ray emission. We consider all the 2FGL sources~\citep{2012ApJS..199...31N} included within a larger region of radius, $R=20^\circ$, to allow for the breadth of the LAT point-spread function that may cause a significant signal from sources outside the ROI to leak into it. The positions and spectral parameters of all 2FGL sources within $15^\circ<R<20^\circ$ from the center of the ROI are fixed to the values reported in the 2FGL catalog; those are on average 21 sources. For the sources within $5^\circ<R<15^\circ$ with $>15\sigma$ detection significance in 2FGL only the flux normalization is let free to vary and all the other parameters are fixed to the values reported in the 2FGL catalog. The parameters for all the other sources within $5^\circ<R<15^\circ$ are fixed to the 2FGL catalog values. Finally, for sources within $R < 5^\circ$ all parameters (index and normalization in case of a power-law spectrum, index, cutoff and normalization in case of a power-law with exponential cutoff and normalization, spectral slope and curvature in case of a log-parabola source spectrum) are free to vary if the source significance exceeds $4\,\sigma$, otherwise all source parameters are fixed. On average 3 sources per ROI have all parameters free, while 6 sources have a free normalization and 18 sources are fixed to the 2FGL values. 

We determine the best values for all the free parameters fitting our source model together with a template for the isotropic and Galactic interstellar emission\footnote{We use the templates provided by the FSSC for the P7REP\_SOURCE\_V15 event class (\url{http://fermi.gsfc.nasa.gov/ssc/data/access/lat/BackgroundModels.html}) with free normalization and free index in case of the Galactic interstellar emission model.} to the LAT data with a binned likelihood approach as described in~\citet{2009ApJS..183...46A}. To quantify the significance of a potential excess above the background, we employ the likelihood-ratio test ~\citep{Neyman1928}.  We form a test statistic 
\begin{equation}
TS=-2\Delta \log{\mathcal{L}} = -2 (\log{\mathcal{L}_0} - \log{\mathcal{L}} ),
\end{equation}
where $\mathcal{L}_0$ is the likelihood evaluated at the best-fit parameters under a background-only, null hypothesis, i.e.~a model that does not include a point source at the SN position, and $\mathcal{L}$ the likelihood evaluated at the best-fit model parameters when including a candidate point source at the SN position. 

\begin{figure}[htbp]
\begin{center}
\includegraphics[scale=\twopic]{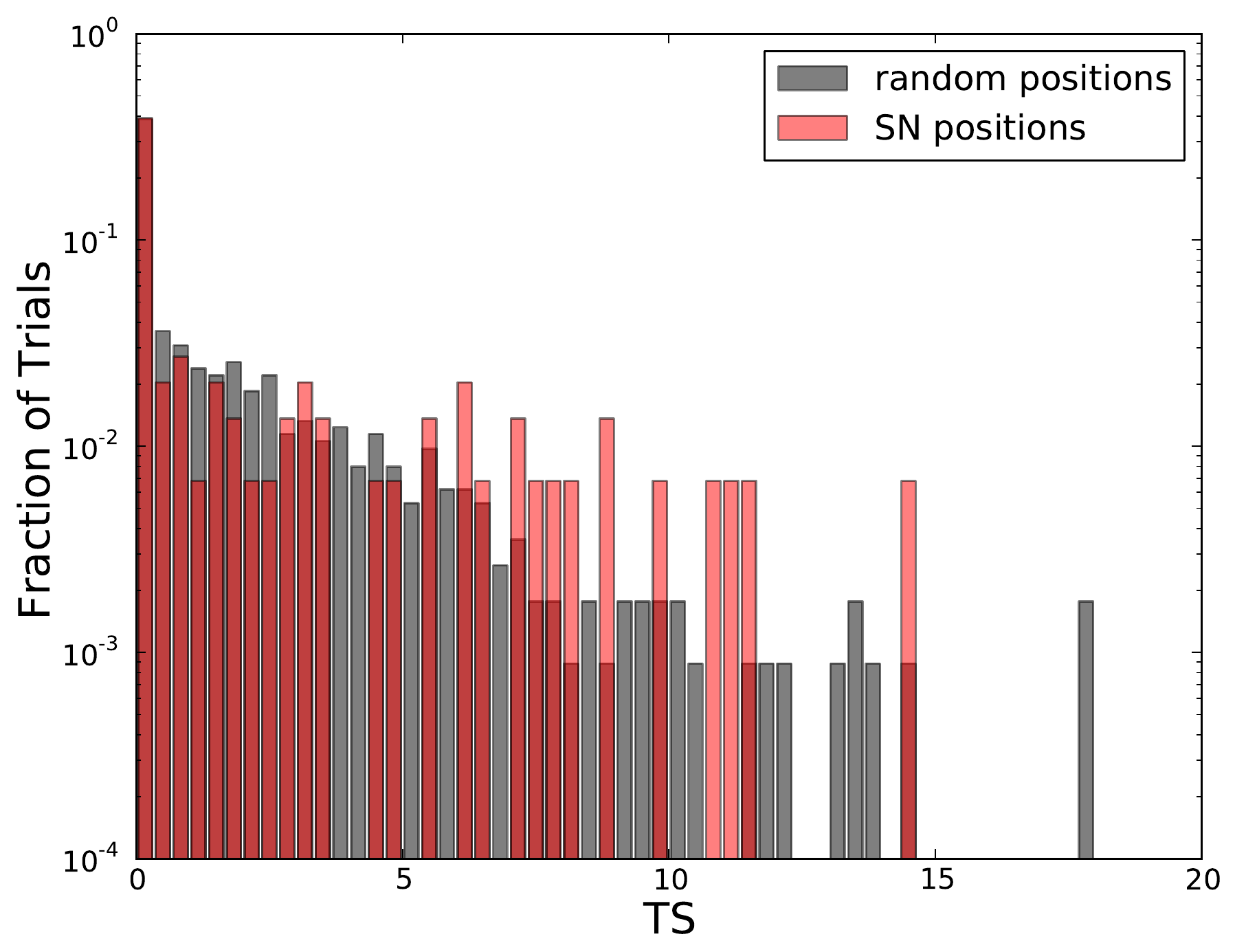}
\includegraphics[scale=\twopic]{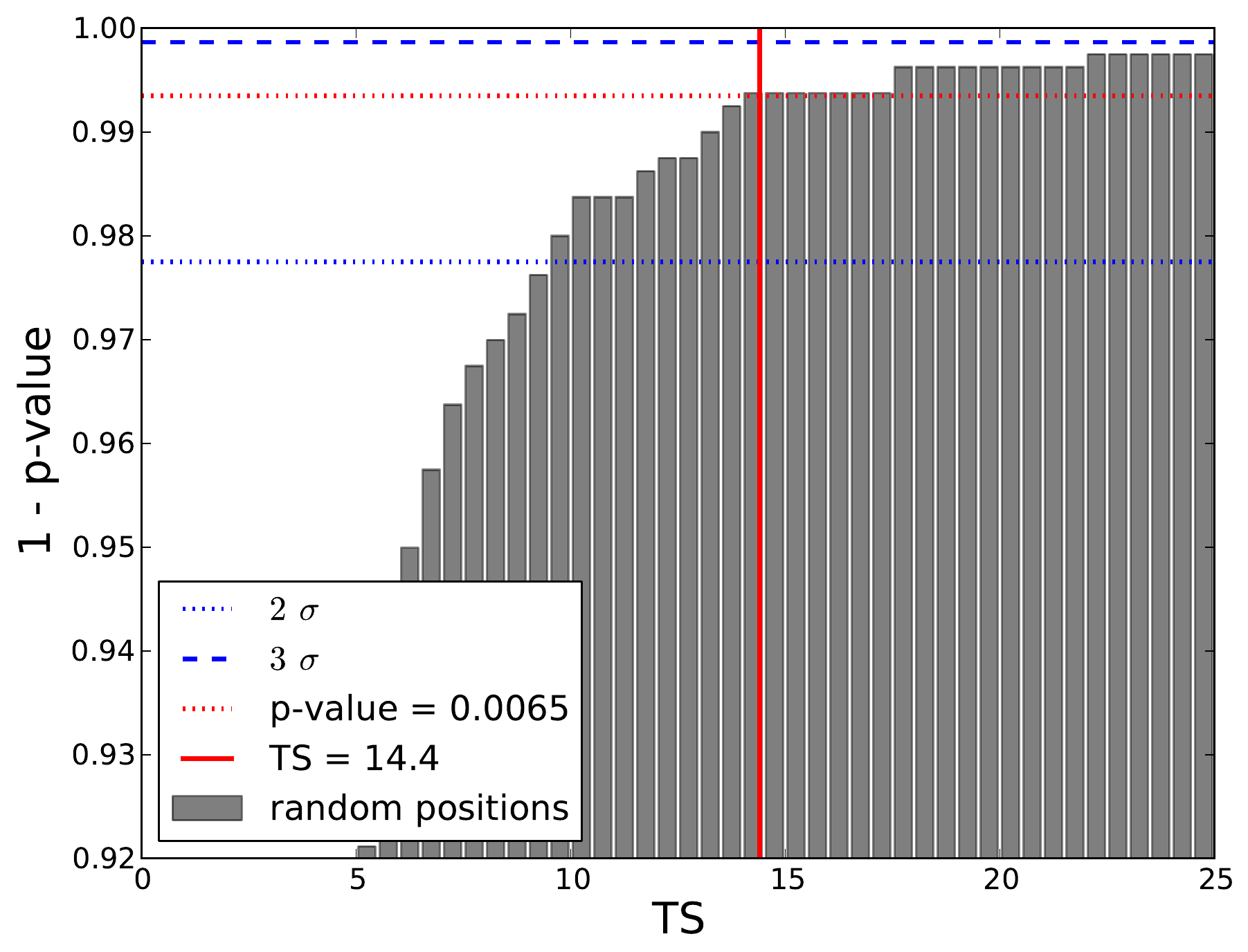}
\noindent
\caption{\small 
Left: Distribution of TS values for a test source modeled by a power-law energy spectrum located at a SN position (red), compared to TS for a similar test source located at a  random position (grey). Right: Cumulative distribution of random-position TS values. The blue dashed (dotted) line indicates a Gaussian equivalent one-sided $3\,\sigma$ ($2\,\sigma$) probability of finding a larger TS than the TS indicated by the intersection of the blue line with the grey distribution. The red solid line shows the largest TS found in the source-specific analysis, which has a p-value of $0.0065$ (red dotted line) and thus lies below $3\,\sigma$. Considering the trials factor, the p-value increases to $0.6$.}
\label{fig:TSHist}
\end{center}
\vspace{1mm}
\end{figure}

The distribution of the TS values obtained for all the SNs using a 1 year time window is displayed in Fig.~\ref{fig:TSHist} (left), compared to the TS distribution obtained from performing a similar analysis at random positions in the sky. We require the random ROI centers to be separated by at least $3.5^\circ$ and to lie outside of the Galactic plane, i.e.~$|b|>10^\circ$. The analysis in the Galactic plane region is complicated by the intense Galactic diffuse emission and none of the SNe in our sample is located close to the plane. Those requirements limit the number of independent ROIs; we use 1140 ROIs in our analysis. The distribution of SN-position TS values is similar to the distribution of random-position TS values (see Fig.~\ref{fig:TSHist} left). The highest TS value found among the SN positions is 14.4, which corresponds to a p-value of $0.0065$ (obtained from the random position analysis), which is below $3\,\sigma$ for a single trial (see Fig.~\ref{fig:TSHist} right). Given the number of SNe in our sample a trials factor needs to be applied, which increases the p-value to $0.6$. 

\begin{figure}[htbp]
\begin{center}
\includegraphics[scale=\twopic]{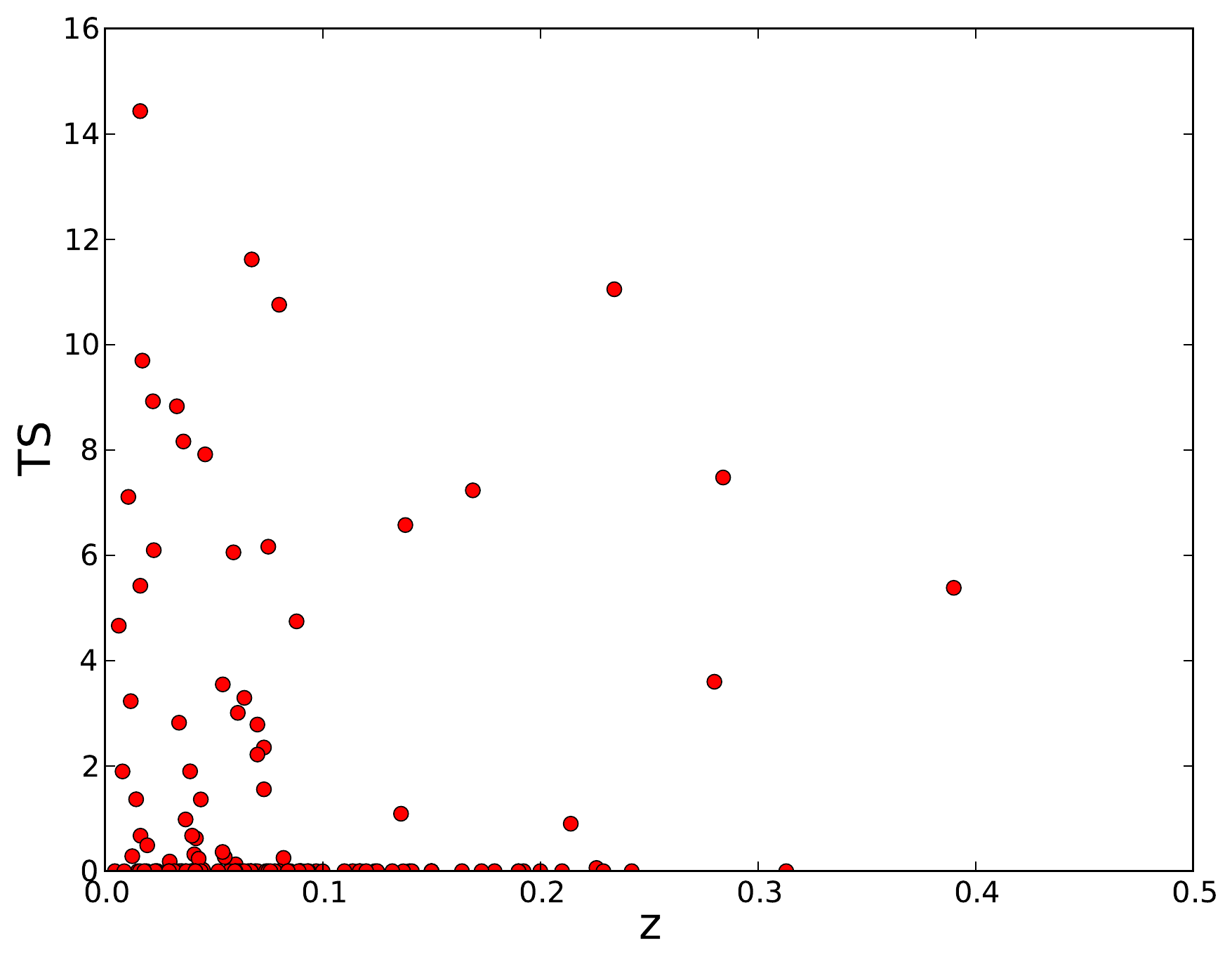}
\includegraphics[scale=\twopic]{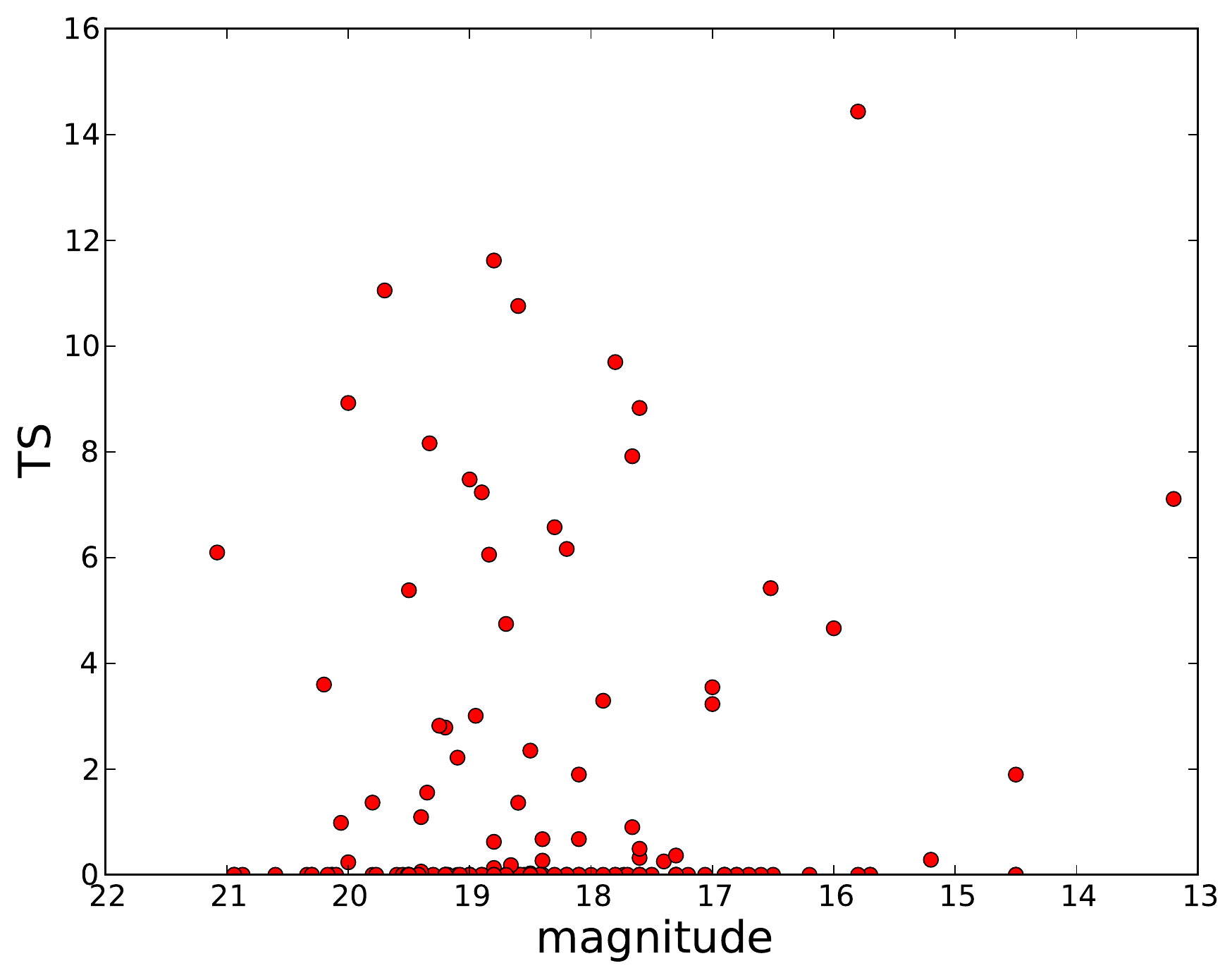}
\noindent
\caption{\small 
TS of a test source located at the SN position as function of redshift (left) and as function of magnitude (right). Note that the largest TS value was $14.4$ (corresponding to a p-value of $0.009$), which does not exceed the detection threshold of $5\,\sigma$.}
\label{fig:TSHistDist}
\end{center}
\vspace{1mm}
\end{figure}

Optically bright SNe are expected to produce a brighter $\gamma$-ray signal than optically dim ones and nearby SNe are expected to be brighter than sources at large distance. However, we do not find an obvious correlation of TS value with redshift or magnitude (see left and right panels of Fig.~\ref{fig:TSHistDist}, respectively), indicating that the $\gamma$-ray signals of individual SNe, if present, are weak.

Three of 147 SNe have a 2FGL source in their close vicinity with an angular distance of less than $0.4^\circ$. In each case the nearest 2FGL source is associated with an active galactic nucleus through multi-wavelength data. Since the spectral parameters of the nearby source are left free to vary in the fit, a possible SNe flux could have been absorbed by the background source. Those sources are PTF10weh, LSQ12by and SN\,2012bq, which are optically dim and distant sources and thus not part of the subsample of nearby and/or bright SNe.

\subsection{Model independent analysis of nearby and/or bright SNe}
\label{subsec:closeby}
\begin{figure}[htbp]
\begin{center}
\includegraphics[scale=\onepic]{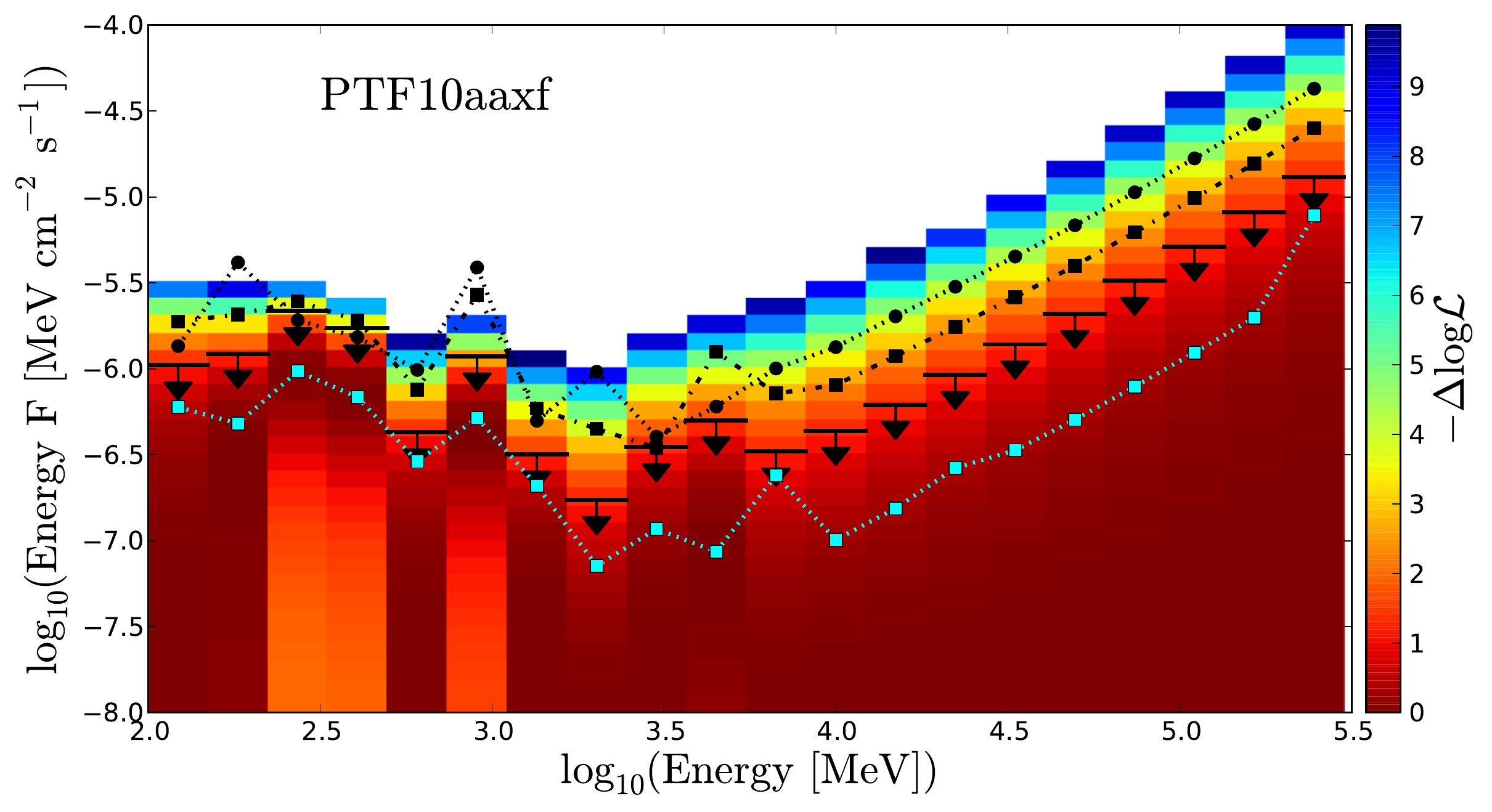}
\noindent
\caption{\small 
Histogram of the bin-by-bin LAT likelihood function used to test for a putative $\gamma$-ray source at the position of supernova SN\,2010jl (PTF10aaxf). The bin-by-bin likelihood is
calculated by scanning the integrated energy flux of the SN within each energy bin
(equivalent to scanning in the spectral normalization of the source). When performing this scan,
the flux normalizations of the background sources are fixed to their optimal values as derived from
a maximum likelihood fit over the full energy range. Within each bin, the color scale denotes the
variation of the logarithm of the likelihood with respect to the best-fit value of the SN
flux using a 1 year time window. Upper limits on the integrated energy flux are set at 95\% CL
within each bin using the delta-log-likelihood technique and are largely independent of the SN spectrum. The black arrows indicate the 95\% CL flux upper limits for $\Delta T = 1$\,yr, where the shown log-likelihood decreases by 2.71/2 from its maximum. For completeness we overlay the 95\% CL upper limits for $~\Delta T = 6$\,months and $\Delta T = 3$ represented by dotted-dashed and dotted lines respectively. For the particular case of SN\,2010jl we repeated the analysis for an extended time window spanning 4.5 years. The results are overlaid as cyan dotted line.}
\label{fig:BinLLH}
\end{center}
\vspace{1mm}
\end{figure}

The $\gamma$-ray spectral shape resulting from particle acceleration in the interaction of SN ejecta with a dense CSM is not known a priori. It is determined by the initial proton spectrum and could be altered by absorption of the $\gamma$ rays in the surrounding medium. Therefore, we study the closest and/or optically brightest sources, which are the most promising sources in terms of expected $\gamma$-ray emission, in an approach independent of a SN spectral model assumption. 
The sources chosen for this analysis have to fulfill the criteria of $z<0.015$ or $m<16.5$, and are listed in Table~\ref{tab:closeSNshort}. 
We fix the spectral parameters of the background sources and the diffuse templates to their global values obtained from the source-by-source analysis over the entire energy range described in Section~\ref{sourceBySource} (without including the SN itself). 
Following the procedure described in~\citet{Ackermann:2013yva} we calculate the likelihood in each of the 20 energy bins after inserting a test source at the SN position at various flux normalization values: 
\begin{equation}
\mathcal{L} (\{\boldsymbol{\mu}_j\}, \boldsymbol{\hat{\theta}} | \mathcal{D}) =   \prod_j  \mathcal{L}_j (\boldsymbol{\mu}_j, \boldsymbol{\hat{\theta}} | \mathcal{D}_j),
\end{equation}
where $\mathcal{D}_j$ is the photon data, $\mathcal{L}_j$ the Poisson likelihood and $\{\boldsymbol{\mu}_j\}$ a set of independent signal parameters in energy bin $j$. The symbol $\boldsymbol{\theta}$ represents the nuisance parameter (i.e., free parameters of background sources and diffuse templates) and $\boldsymbol{\hat{\theta}}$ indicates that they have been fixed to their global values. 
The bin-by-bin likelihood allows us to find the upper limits at $95\%$ CL\footnote{Note, we are using a two-sided confidence interval.}, defined as the value of the energy flux, where the log-likelihood decreases by $2.71/2$ from its maximum~\citep[the ``delta-log-likelihood technique" --][]{Bartlett1953,2005NIMPA.551..493R}.  
An example is shown in Fig.~\ref{fig:BinLLH} for SN\,2010jl~\citep[PTF10aaxf --][]{2014ApJ...781...42O,2014ApJ...797..118F,Zhang2012}, while similar plots for all nearby sources can be found in Appendix~\ref{sec:LLHProfiles}. Any SN model predicting a certain $\gamma$-ray spectrum can be tested using those results~\citep[see][for more details on the bin-by-bin likelihood]{Ackermann:2013yva} by recreating a global likelihood by tying together the signal parameters over the energy bins:
\begin{equation}
\mathcal{L} ({\boldsymbol{\mu}}, \boldsymbol{\hat{\theta}} | \mathcal{D}) =   \prod_j  \mathcal{L}_j ({\boldsymbol{\mu}_j (\boldsymbol{\mu})},  \boldsymbol{\hat{\theta}}| \mathcal{D}_j),
\label{eq:globalLLH}
\end{equation}
with ${\boldsymbol{\mu}}$ denoting the global signal parameters.

For the most promising source of our sample, SN\,2010jl, we repeat the analysis for an extended time window ending in May 2015, i.e., spanning 4.5 years. This is motivated by the fact that in some cases SN Type IIn emission lasts for 3-5 years after the explosion~\citep{2009Natur.460..237C}.

\subsection{Joint Likelihood Analysis}
\label{subsec:stacking}
For greater sensitivity to a weak $\gamma$-ray signal from interaction-powered SNe, we combine the 16 closest and/or brightest sources in a joint likelihood analysis. To be independent from any spectral shape assumption we perform the analysis in energy bins (see Section~\ref{subsec:closeby} for details of the bin-by-bin likelihood analysis). In each energy bin we tie the SN flux normalization for all 16 SNe together resulting in one free parameter per energy bin. The likelihood values for the individual sources, $i$, are multiplied to form the joint likelihood:  
\begin{equation}
\mathcal{L}(\boldsymbol{\mu},\{\boldsymbol{\hat{\theta}}_i\}| \mathcal{D}) = \prod_i \mathcal{L}_i (\boldsymbol{\mu}, \boldsymbol{\hat{\theta}}_i | \mathcal{D}_i).
\end{equation}
However, we have to make some assumption about a common scaling factor of the $\gamma$-ray flux in order to tie the SNe flux normalizations together (i.e., we want to give a larger weight to SNe with greater expected $\gamma$-ray fluxes in the joint likelihood). We use two different approaches: 
first, we assume that all SNe have the same intrinsic $\gamma$-ray luminosity; therefore, the observed $\gamma$-ray flux for each SN scales with a factor inversely proportional to the square of the luminosity-distance $d$. The redshift is measured for each SNe and since we only consider nearby SNe we use a simple linear approximation for the relation between redshift and distance: $d = z \times c/H$, with $H = 67.8$\,km\,s$^{-1}$\,Mpc$^{-1}$~\citep{2013arXiv1303.5062P}. We do not apply a redshift-dependent energy rescaling for SNe at different redshifts, since the energy shift is negligible at the small redshifts (i.e., $z<0.015$) considered in this analysis. We weight the flux normalization in each energy bin of each source with $w_d = (10\,\textrm{Mpc}/d)^2$. We then tie those weighted normalizations together. The exact value of $H$ does not influence our results since the combined normalization of all sources is free in the fit of the model to the data in each energy bin. Note that only the SN flux normalization is free while the background source parameters as well as the diffuse template parameters are fixed to their global values obtained from a fit to the entire energy range.

Alternatively, we assume that the $\gamma$-ray flux is correlated with the optical flux, i.e.~we use a weight proportional to the optical flux\footnote{Note that flux and apparent magnitude are related through: $m-m_0 = -2.5 \log_{10} \frac{F}{F_0}$, where $F_0$ and $m_0$ are the flux and apparent magnitude of a reference star.} or $10^{-0.4m}$. We chose the weight to be: 
\begin{equation}
w_m = 10^{-0.4(m-C)} = 10^{-0.4m+5.2},
\end{equation}
where $m$ is the apparent $R$-band magnitude provided by the SN catalog and $C=13$ is a normalization constant. 
Again, the exact choice of $C$ does not influence our results since the combined normalization of all sources is free in the fit. We chose to neglect a correction for Galactic dust extinction, which is at most $0.28$\,mag and thus smaller than the uncertainty in the peak magnitude determination.

\begin{figure}[htbp]
\begin{center}
\includegraphics[scale=\twopic]{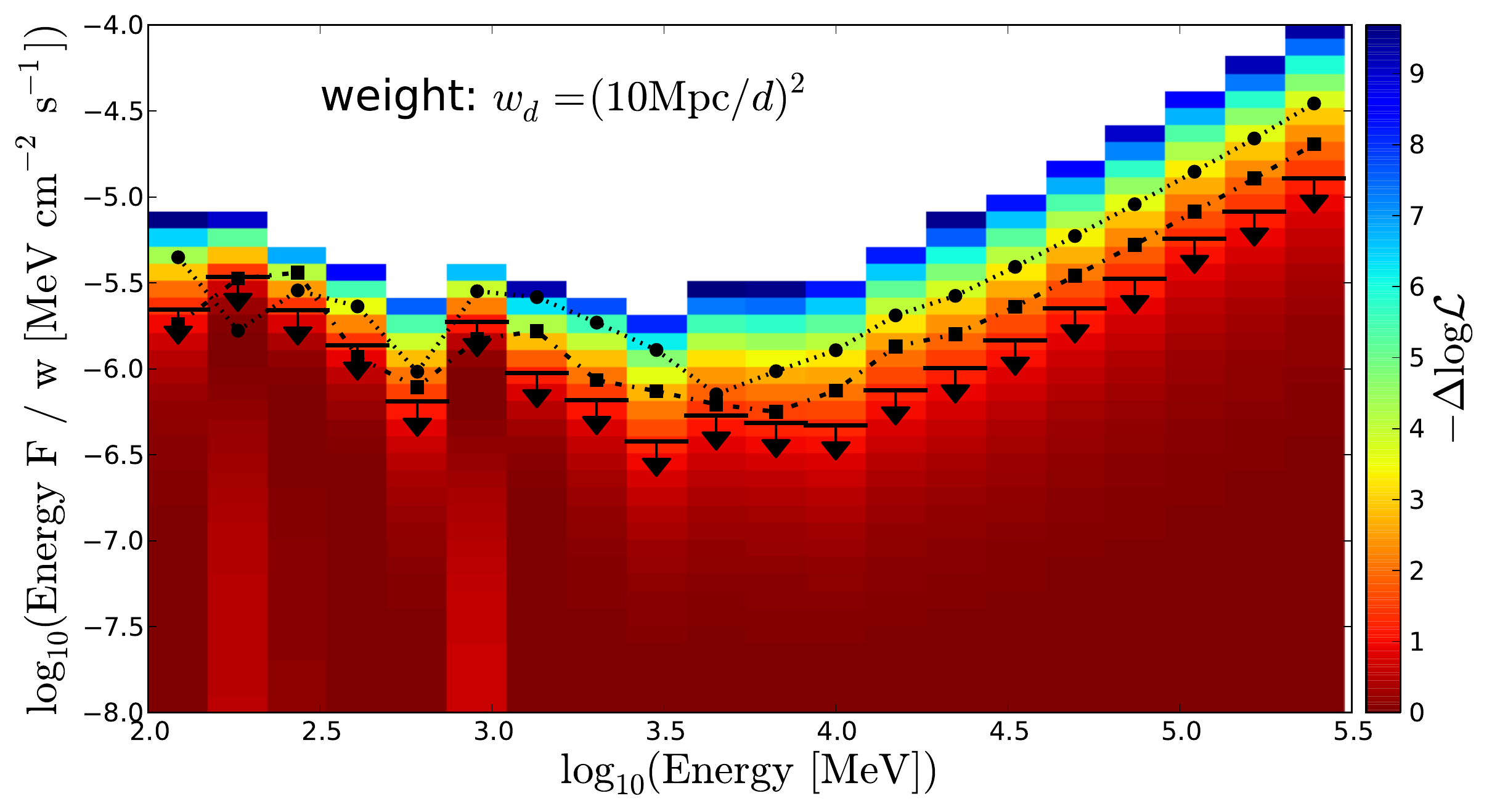}
\includegraphics[scale=\twopic]{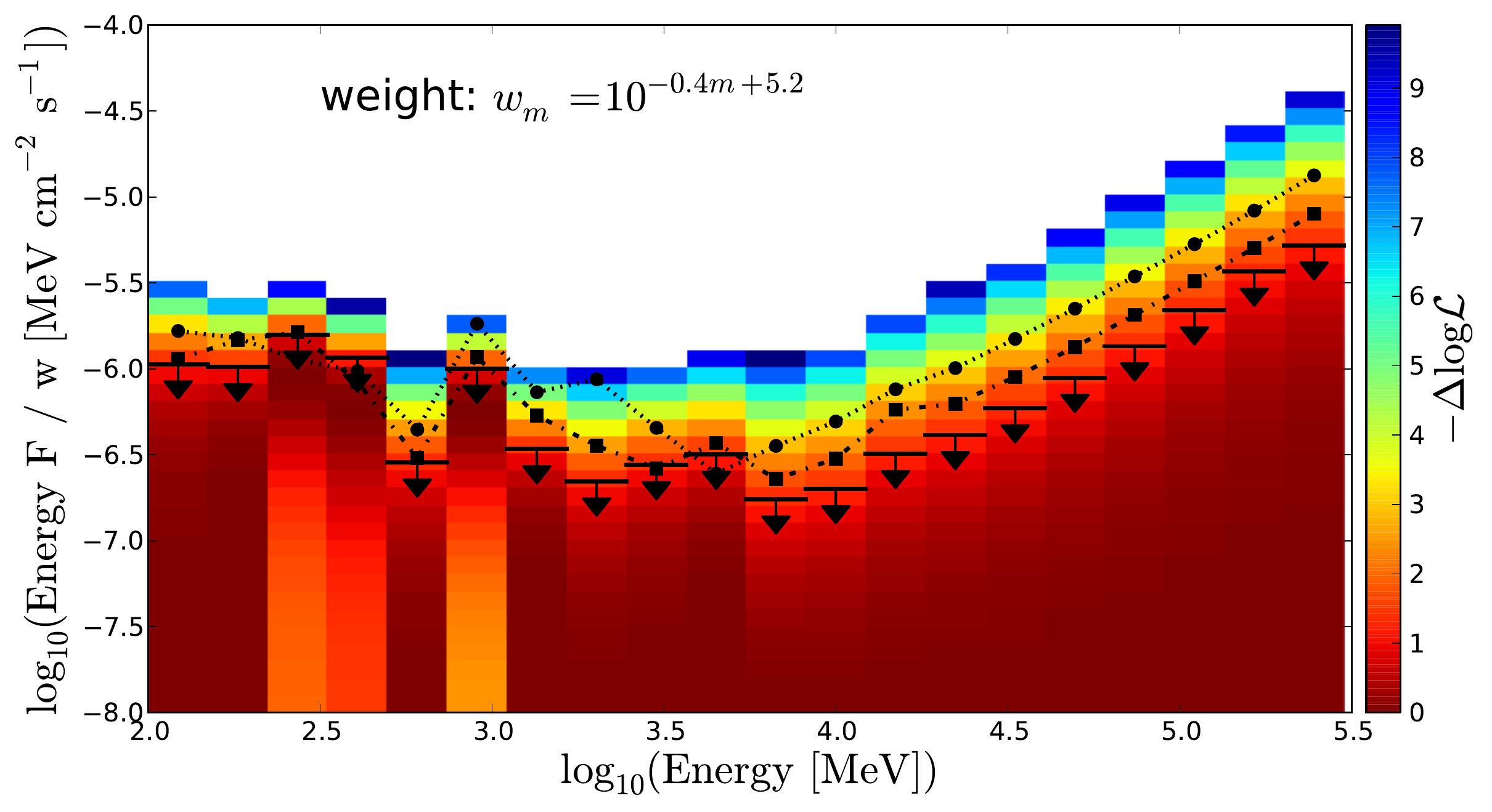}
\noindent
\caption{\small 
Similar to Fig.~\ref{fig:BinLLH}, but for the composite likelihood instead of the single-source likelihood.
Left: Composite likelihood profile for each energy bin weighting each source with $(10$\,Mpc$/d)^2$. Right: Composite likelihood profile for each energy bin weighting each source with $10^{-0.4m+5.2}$. The black arrows indicate the 95\% upper limits for $\Delta T = 1$\,yr, while the dotted-dashed and dotted lines represent the 95\% upper limits for $~\Delta T = 6$\,months and $\Delta T = 3$\,months, respectively.}
\label{fig:compLLH}
\end{center}
\vspace{1mm}
\end{figure}

We perform the joint likelihood analysis for three time windows: 1\,year, 6\,months, 3\,months since the $R$-band maximum light. Fig.~\ref{fig:compLLH} shows the likelihood profiles of the combined $\gamma$-ray flux. Table~\ref{tab:LLHSummary} summarizes the results from the combined likelihood analysis and shows the sum of TS over all energy bins.
No significant improvement in the likelihood by including the SNe in the fit could be found in the joint likelihood analysis. The largest TS value of $8.8$ is found in case of assuming the $\gamma$-ray flux scales with the optical flux for the 1-year time windows. According to Wilks' theorem, TS is distributed approximately as $\chi^{2}$ with the degrees of freedom equal to the number of parameters characterizing the additional source.
Taking into account the number of free parameters (20, one for each energy bin) the probability that this is a statistical fluctuation is $98.5\%$. This significance would be further decreased by taking into account trials factors for the two different weighting schemes and 3 different time windows.

However, if we assume a spectral model for the SN flux, we can greatly reduce the number of free parameters. For illustration we fit a power-law spectral shape to the bin-by-bin likelihood following Equation~\ref{eq:globalLLH}. The index and normalization of the power-law function are left free to vary in the fit. The resulting TS values and corresponding p-values (not including trials factors) are summarized in Table~\ref{tab:LLHSummary}; none of them are significant. A more physical spectral model is fitted to the bin-by-bin likelihood in Section~\ref{sec:Interpretation}.

\begin{center}
\begin{table}[htb]
{\footnotesize
  \begin{tabular}{ l  c c c c c c}
    \hline
    Weighting & \multicolumn{3}{c}{TS} & \multicolumn{3}{c}{TS$_{PL}$ (p-value)}\\
     & 1\,yr & 6\,months & 3\,months & 1\,yr & 6\,months & 3\,months\\
    \hline
    $(\rm{10\,Mpc}/d)^2$ & 2.2 & 2.1 & 2.4 & 0.0 (1.0) & 0.0 (1.0) & 0.0 (1.0)\\
    $10^{-0.4m+5.2}$ & 11.7 & 7.8 & 9.0 & 2.9 (0.23) &1.6 (0.45) & 0.0 (1.0) \\
    \hline
  \end{tabular}
  }
  \caption{Sum over bin-by-bin TS values obtained from the joint likelihood analysis. TS$_{PL}$ is the TS obtained by assuming a power-law spectral shape.}
    \label{tab:LLHSummary}
  \end{table}
\end{center}

\subsection{Joint Likelihood Analysis of SN Subsample with Confirmed Massive CSM}
\begin{figure}[htbp]
\begin{center}
\includegraphics[scale=\onepic]{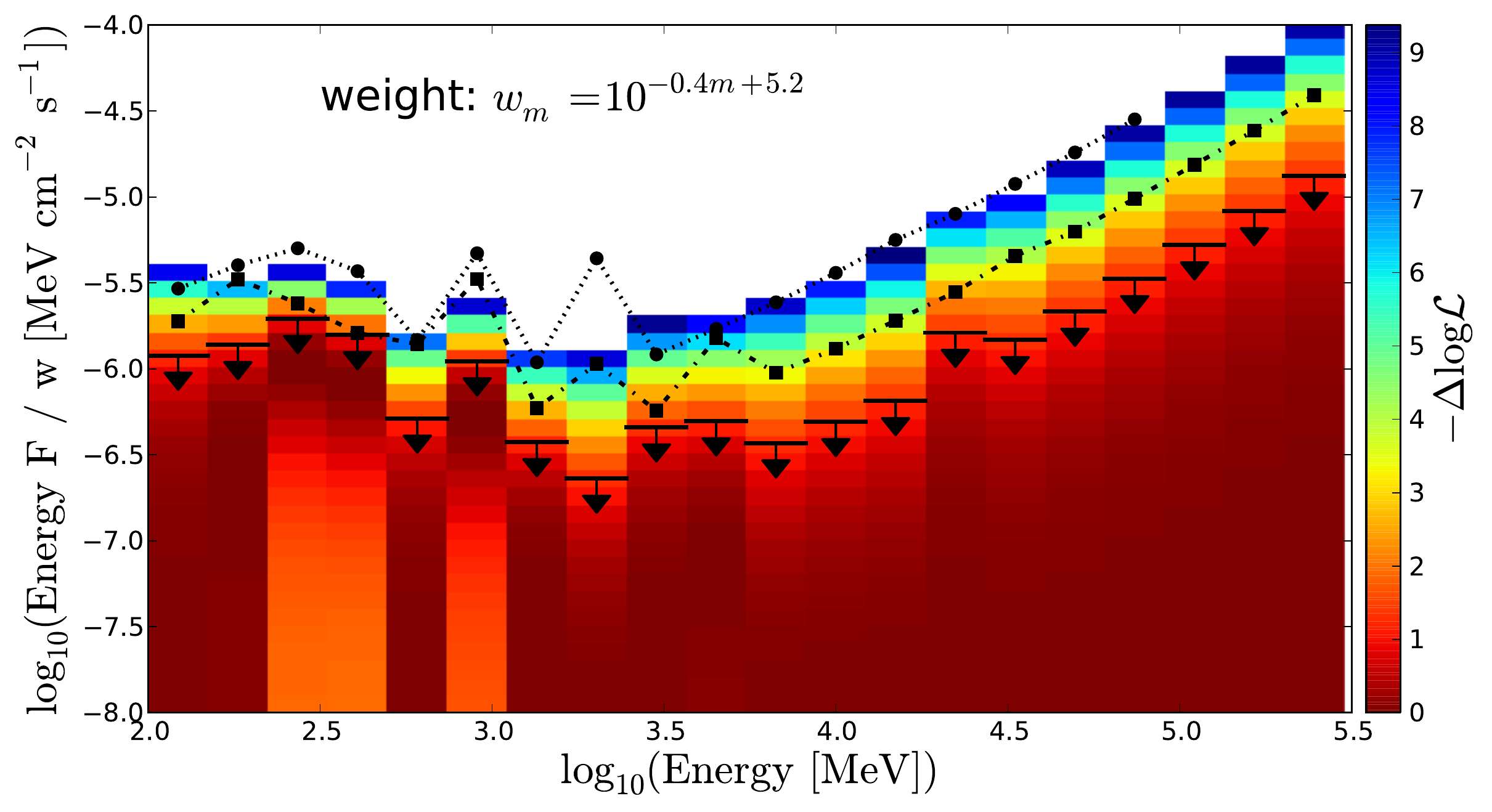}
\noindent
\caption{\small 
Joint likelihood analysis of the SN subsample with confirmed massive CSM: Joint likelihood profile for each energy bin weighting each source with $10^{-0.4m+5.2}$. The black arrows indicate the 95\% upper limits for $\Delta T = 1$\,year, while the dotted-dashed and dotted lines represent the 95\% upper limits for $~\Delta T = 6$\,months and $\Delta T = 3$\,months, respectively.}
\label{fig:eranStacking}
\end{center}
\vspace{1mm}
\end{figure}

We select a subsample of 16 SNe from the Type IIn SNe catalog for which we have additional evidence through multi-wavelength observations for the existence of a massive CSM.
We select SNe that show Balmer emission lines and continuum in both early and late times. The SNe in this sample are: PTF12csy, PTF11oxu, PTF11mhr, PTF11fzz, PTF11fuu, PTF10aaxf, PTF10ptz, PTF10scc, PTF10jop, PTF10fei, PTF10qaf, PTF10tel, PTF10tyd, PTF10gvf, PTF10cwl, PTF09drs. We repeat the joint likelihood analysis described above for this subset with the optical flux weighting scheme for three time windows (1 year, 6 months and 3 months). The results are displayed in Fig.~\ref{fig:eranStacking}. The TS values of the composite fit are 11.3, 17.5 and 10.3 for the time windows of  1 year, 6 months, and 3 months, respectively. Taking into account the 20 free parameters, the chance probability for a TS of 17.5 is $62\%$.

\section{Interpretation}
\label{sec:Interpretation}

\citet{Murase:2010cu} suggested that $\gamma$-ray emission is produced by cosmic rays accelerated at the early collisionless shock between SN ejecta and circumstellar material. 
For the scenario described by~\citet{Murase:2013kda}, $\gamma$-ray emission can be predicted when the model parameters are determined by optical and X-ray observations. We defer such  model-dependent analyses to future work. Instead, in this work, we take a model-independent approach, where we aim to constrain the $\gamma$-ray luminosity as a function of the proton spectral index. We assume that the spectrum of CR protons is given by a power law (in momentum) with the minimum and maximum proton momenta of $0.1$\,GeV/c and ${10}^8$\,GeV/c, respectively. Then, we calculate the $\gamma$-ray flux following~\citet{2006PhRvD..74c4018K}.  In the calorimetric limit, which is expected for SNe like SN 2010jl~\citep{Murase:2013kda}, the $\gamma$-ray spectral index follows the proton spectral index, although the resulting limits (shown in Fig.~\ref{fig:lumLimits}) are similar to what would be obtained for non-calorimetric cases, for which the resulting shape of the $\gamma$-ray spectrum is slightly harder than the proton spectral shape due to the energy dependence of the $pp$ cross section.  For simplicity we do not take into account $\gamma$-ray absorption; \citet{Murase:2013kda} showed that GeV $\gamma$ rays can escape from the system without severe matter attenuation if the shock velocity is high enough.  

The diffusive shock acceleration theory predicts that the proton acceleration efficiency is $\epsilon_p\sim0.1$. In the calorimetric limit, all the proton energy is used for pion production, and $1/3$ of pions are neutral pions that decay into $\gamma$ rays. Then, about half of the $\gamma$ rays are absorbed deep inside the ejecta, so we expect $L_\gamma\approx(1/6)\epsilon_pf_{\rm esc}L_{\rm kin}$, where $L_{\rm kin}$ is the kinetic luminosity and $f_{\rm esc}$ is the escape fraction of $\gamma$ rays. The $\gamma$-ray attenuation due to the Bethe-Heitler process is relevant when the shock velocity is lower than $\sim4500~{\rm km}~{\rm s}^{-1}$, while the two-photon annihilation process is relevant when the shock velocity is high enough~\citep{Murase:2013kda}.  Although $\gamma$ rays can escape late after the shock breakout, the attenuation can be relevant around the shock breakout so we assume $f_{\rm esc}\sim0.1$-$1$ to take into account uncertainty of the $\gamma$-ray flux.  
The radiation energy fraction is given by $\epsilon_\gamma\equiv L_{\rm rad}/L_{\rm kin}$, where $L_{\rm rad}$ is the bolometric radiation luminosity.  About half of the kinetic energy is converted into the thermal energy, and half of the thermal energy is released as outgoing radiation, which implies $\epsilon_\gamma\sim1/4$~\citep{2014ApJ...781...42O}.  As a result, we have $L_\gamma/L_{\rm rad}\approx(1/6)(\epsilon_p/\epsilon_\gamma)f_{\rm esc}\sim(1/15)f_{\rm esc}$. Our limits presented below are on the fraction of $\gamma$-ray to R-band luminosity, which is an upper bound on $L_\gamma/L_{\rm rad}$. In the case of SN\,2010jl  $L_{R}\sim L_{\rm rad}$ and thus $L_\gamma/L_R\sim0.01$-$0.1$ is theoretically expected.

\begin{figure}[htbp]
\begin{center}
\includegraphics[scale=\twopic]{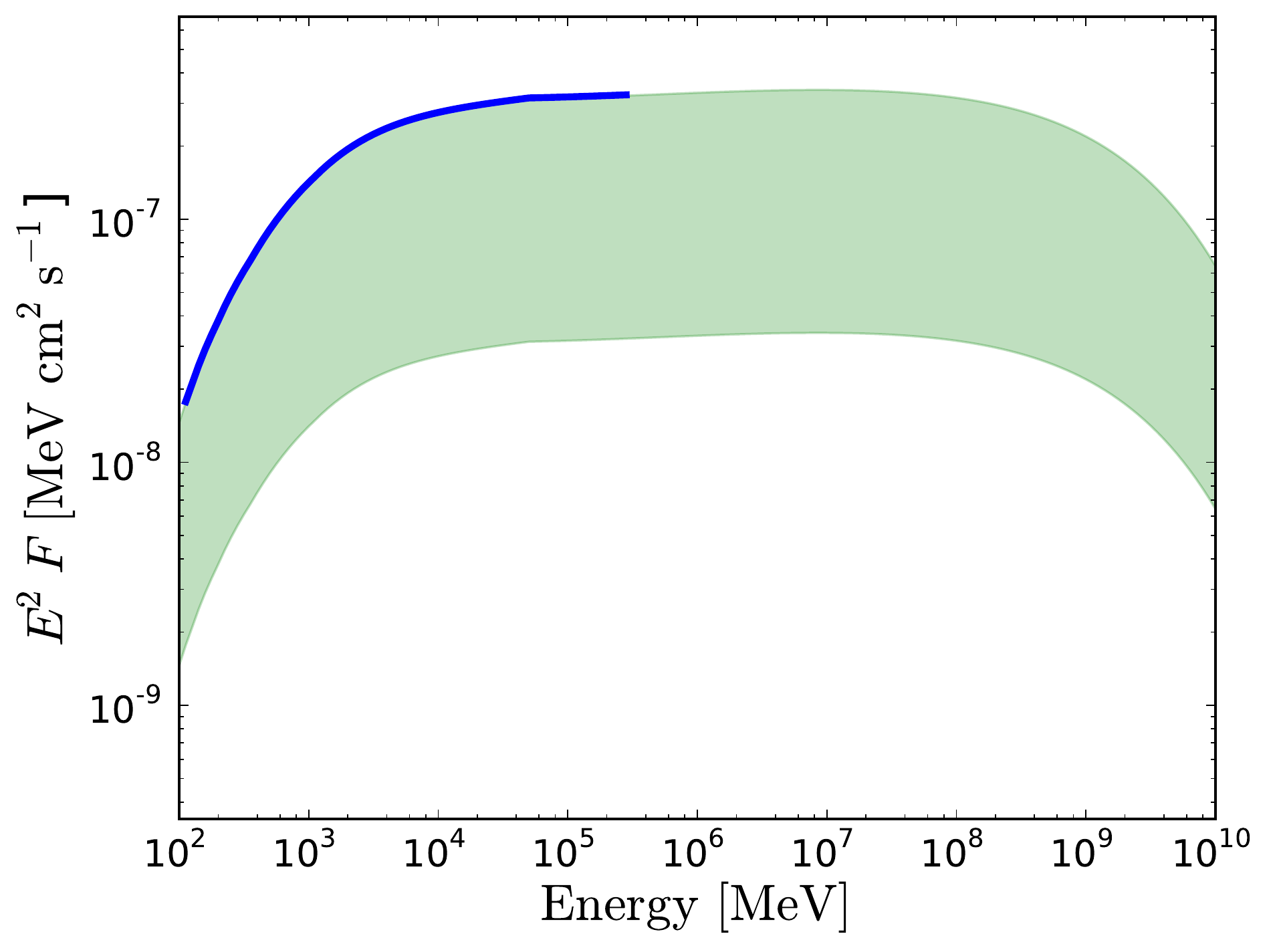}
\includegraphics[scale=\twopic]{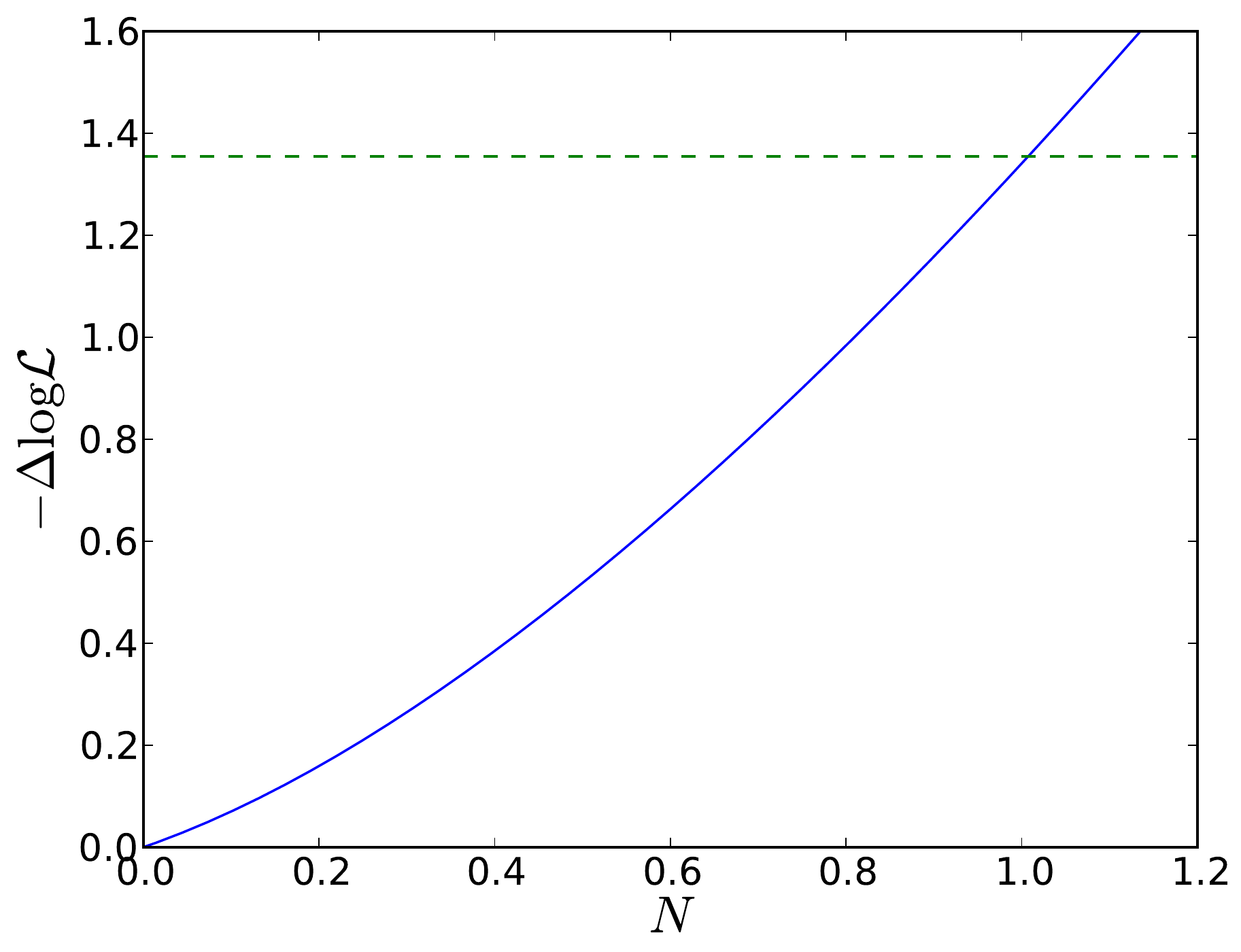}
\noindent
\caption{\small 
Left: Predicted $\gamma$-ray energy spectrum for SN\,2010jl assuming $\Gamma_p = -2$ and a normalization of the $\gamma$-ray flux yielding $0.01 < L_{\gamma}/L_{R}  < 0.1$ shown as the green shaded region compared to the $95\%$ flux upper limit (blue). Right: Likelihood profile for the spectral normalization parameter $N$ relative to the flux prediction yielding $L_{\gamma}/L_{R} = 0.1$. The dashed green line indicates an increase of the negative delta log-likelihood by 2.71/2 compared to its minimum.
}
\label{fig:SN2010jlFlux}
\end{center}
\vspace{1mm}
\end{figure}

As an example, we consider supernova SN\,2010jl (PTF10aaxf), which is the most-likely detectable CR accelerator, because multi-wavelength observations indicate a very massive CSM of 10\,M$_{\odot}$.  We present a generic flux prediction for the calorimetric limit for this source assuming a proton spectral index of $\Gamma_p = - 2$ and a normalization of the $\gamma$-ray flux that yields $0.01<  L_{\gamma}/L_{R}  < 0.1$ (shown as shaded green region in Fig.~\ref{fig:SN2010jlFlux}) and calculate the corresponding flux upper limit (shown in blue in Fig.~\ref{fig:SN2010jlFlux}) following the procedure outlined in~\cite{Ackermann:2013yva}.  
The bin-by-bin likelihood analysis is used to re-create a global likelihood for a given signal spectrum by tying the signal parameters across the energy bins (see Eqn.~\ref{eq:globalLLH}). In this case the global signal parameter is the flux scale factor $N$ relative to the flux that yields  $L_{\gamma}/L_{R} =0.1$ (i.e., the upper bound of the uncertainty band shown in Fig.~\ref{fig:SN2010jlFlux}, left). 
We assume that SN\,2010jl is at distance 48.7 Mpc with an apparent R-band peak magnitude of 13.2. 
We calculate the change in log-likelihood for various values of $N$ and find the $95\%$ flux upper limit (given by the value of $N$ for which the delta log-likelihood decreases by 2.71/2 compared to its minimum). The derived upper limit touches the optimistic model prediction, i.e. the upper bound of the theoretical uncertainty band. A more detailed modeling of the expected flux based on multi-wavelength observations is outside the scope of this paper and will follow in future work. Better constraints on the $\gamma$-ray escape fraction are crucial to calculate stringent limits on the proton acceleration efficiency and will be obtained in more detailed modeling.

\begin{figure}[htbp]
\begin{center}
\includegraphics[scale=\onepic]{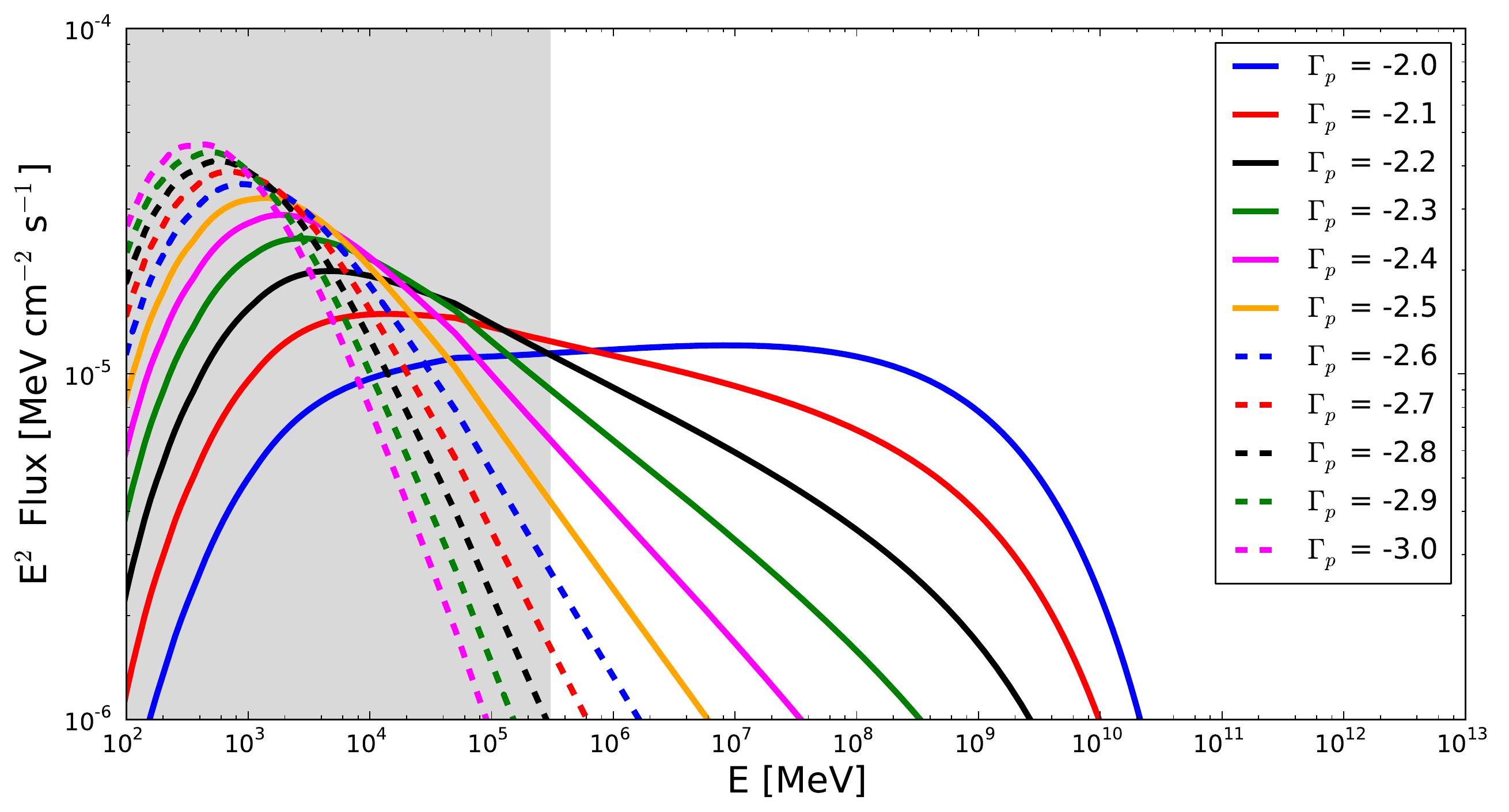}
\noindent
\caption{\small 
Gamma-ray energy spectra assuming a total $\gamma$-ray energy of $10^{50}$\,erg, a source distance of $10$\,Mpc and a duration of 1 year for various proton spectral indices $\Gamma_p$. The shaded grey region shows the energy range covered by this analysis.
}
\label{fig:gammaSpecIP}
\end{center}
\vspace{1mm}
\end{figure}

\begin{figure}[htbp]
\begin{center}
\includegraphics[scale=\twopic]{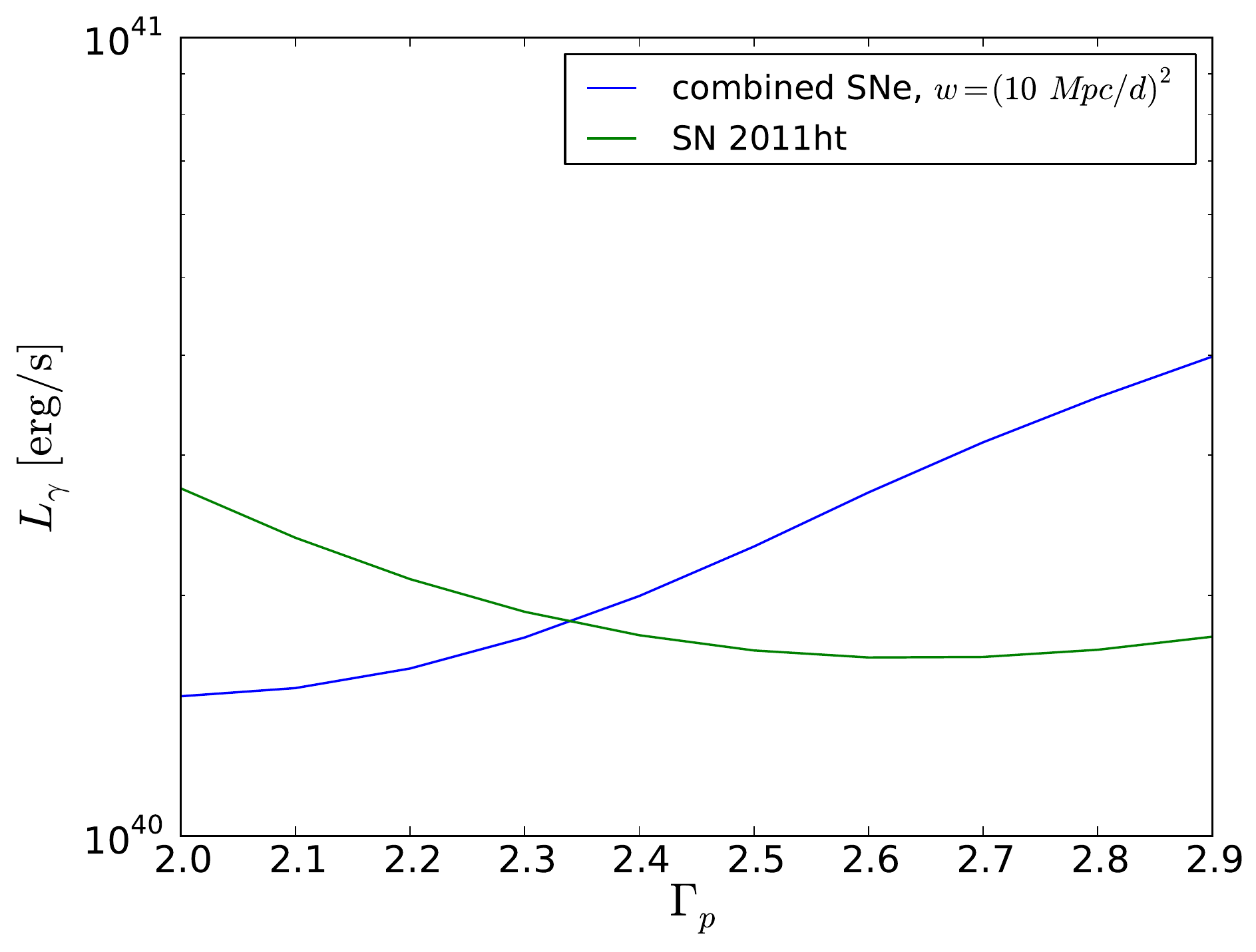}
\includegraphics[scale=\twopic]{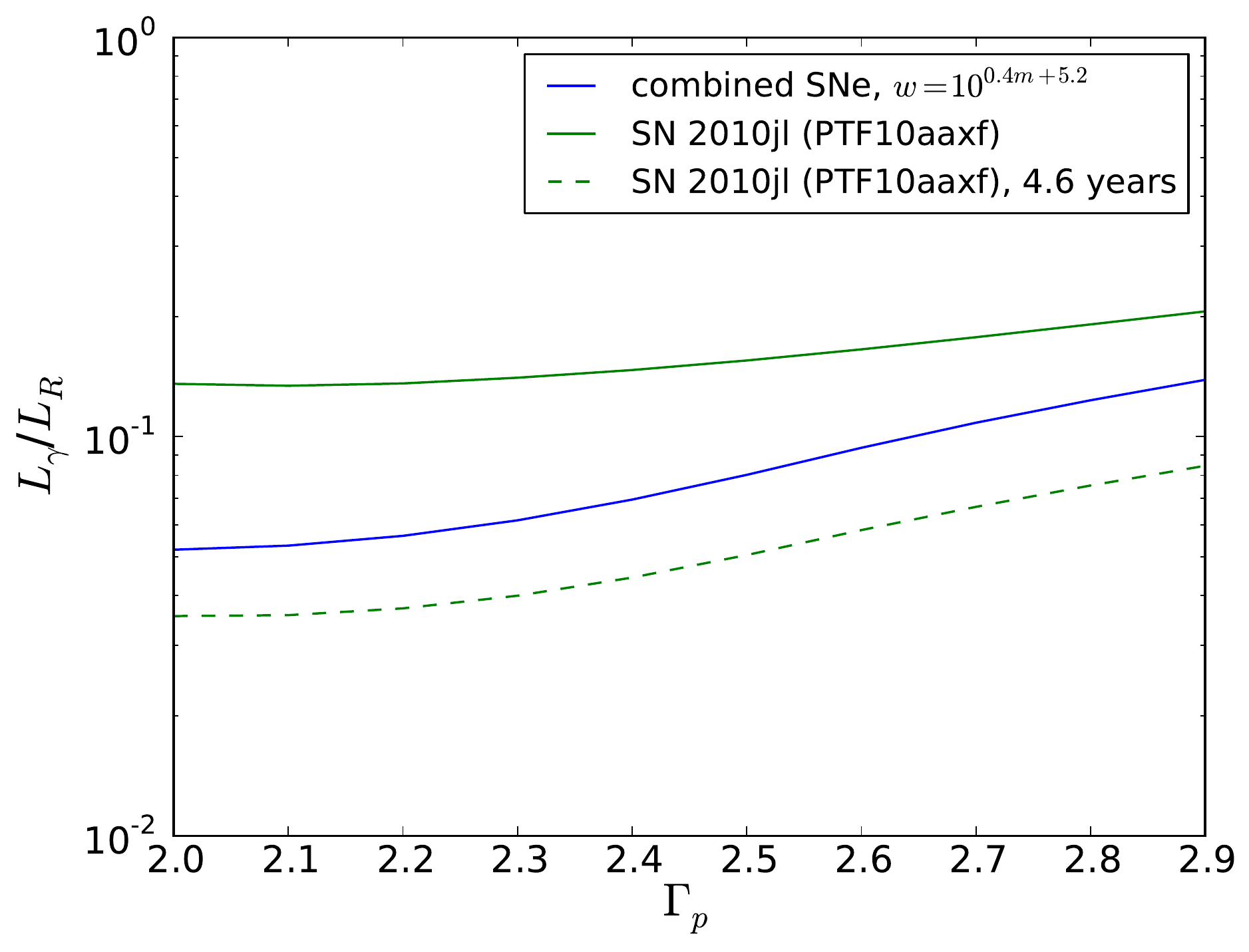}
\noindent
\caption{\small 
Left: $95\%$ CL upper limit on the $\gamma$-ray luminosity as a function of the proton spectral index based on the results obtained from the joint likelihood analysis with $1/d^2$ weighting shown in blue compared to the limit obtained from the closest single source SN\,2011ht in green. Right: 
$95\%$ CL upper limit on the ratio of $\gamma$-ray and optical luminosity $L_{\gamma}/L_{R}$ as a function of $\Gamma_p$ assuming 
a proportionality between optical and $\gamma$-ray flux shown in blue compared to the limit obtained from a single source analysis of SN\,2010jl considering a one year time window (in green). The results of the analysis with an extended time window of 4.5 years for SN\,2010jl are shown in dashed green.
}
\label{fig:lumLimits}
\end{center}
\vspace{1mm}
\end{figure}

More stringent limits are expected from the joint likelihood results\footnote{Note that including sources with a statistical over-fluctuation can worsen the joint limit.}. 
Generic $\gamma$-ray flux predictions for various proton spectral indices are shown in Fig.~\ref{fig:gammaSpecIP}. We calculate the $95\%$ CL upper limit on the $\gamma$-ray luminosity 
\begin{equation}
L_{\gamma} = 4\pi d^2 F_{\gamma}^I = 4\pi (10\,\textrm{Mpc})^2 \frac{F_{\gamma}^I}{w_d},
\end{equation}
where $F_\gamma^I$ is the integrated $\gamma$-ray flux over the energy range used in this analysis. The luminosity $L_{\gamma}$ is proportional to the result of the joint likelihood analysis using the weight $w_d = (10\,\textrm{Mpc}/d)^2$, assuming all sources have the same $L_{\gamma}$.
In other words our joint likelihood results set a limit on $F_{\gamma}^I / w_d$ and thus on $L_{\gamma} $. The result is shown in Fig.~\ref{fig:lumLimits} (left) as a function of the proton spectral index.

In addition we calculate the $95\%$ CL upper limit on the ratio of $\gamma$-ray to optical luminosity 
\begin{equation}
L_{\gamma}/L_{R} = \frac{4\pi d^2 F_{\gamma}^I}{L_\sun 10^{0.4(M_\sun - M)}} =  \frac{4\pi (1\,\textrm{Mpc})^2}{L_\sun 10^{0.4M_\sun+4.8}} \frac{F_{\gamma}^I}{w_m},
\end{equation}
where $L_\sun = 6 \times 10^{32}$\,erg\,s$^{-1}$ is the R-band luminosity and $M_\sun = 4.7$ the absolute $R$-band magnitude of the Sun. The ratio is proportional to $F_\gamma/w_m$, which is constrained by the joint likelihood analysis assuming a correlation of optical and $\gamma$-ray flux, i.e. weighting with $w_m = 10^{-0.4m+5.2}$. Thus we can use the joint likelihood results to set a limit on
$L_{\gamma}/L_{R}$ as a function of $\Gamma_p$ (see Fig.~\ref{fig:lumLimits} right).

In Fig.~\ref{fig:lumLimits} both limits discussed above are compared to the limit obtained using only one SN. The closest SN (SN\,2011ht with a distance of $d=17.7$\,Mpc) in the case of $1/d^2$ weighting and the brightest SN (SN\,2010jl with a magnitude of $m=13.2$) in case of weighting with the optical flux. In both cases the combined limit is dominated by one SN. In the case of $1/d^2$ weighting the single source limit is better than the combined limit, indicating a statistical under-fluctuation in the individual analysis of this source or an over-fluctuation in one of the sources included in the joint likelihood.

\section{Conclusions}
\label{sec:Conclusions}

The origin of the multi-wavelength emission of Type IIn SNe and the onset of cosmic-ray production in supernova remnants is not fully understood. Type IIn SNe are expected to be host sites of particle acceleration, which could be pinpointed by transient $\gamma$-ray signals. 
For the first time we searched in a systematic way for $\gamma$-ray emission from a large ensemble of Type IIn SNe in coincidence with optical signals. No evidence for a signal was found, but our observational limits start to reach interesting parameter ranges expected by the theory. We set stringent limits on the $\gamma$-ray luminosity and the ratio of $\gamma$-ray and optical luminosity. For example, we can exclude $L_\gamma/L_R >0.1$ at $95\%$ CL for proton spectral indices of $<2.7$ from the results of the combined likelihood analysis assuming that $L_\gamma/L_R$ is constant. Those constraints can be converted to limits on the proton acceleration efficiency. In the case of SN\,2010jl, our limits are close to theoretically expected values.  However, uncertainties in the modeling, including the $\gamma$-ray escape fraction, leads to the range of $O(10\%)$ to $O(1\%)$ for the ratio of $\gamma$-ray to optical luminosity. Model-dependent calculations based on multi-wavelength observations will be performed in a future work and will allow us to set stringent constraints on the proton acceleration efficiency.

We do not have to make this assumption in the analysis of individual SNe. The results from the optically brightest SN in our sample, SN\,2010jl, alone leads to only a factor of two weaker constraints, excluding $L_\gamma/L_R >0.2$. Assuming a scaling of the $\gamma$-ray flux with $1/d^2$ we can exclude $L_\gamma>4\times 10^{40}$\,erg\,s$^{-1}$ at $95\%$ CL for all indices considered. A total $\gamma$-ray luminosity of $10^{50}$\,erg emitted within 1\,year (as assumed in Fig.~\ref{fig:gammaSpecIP}) is excluded. The limits presented here are based on minimal assumptions about the $\gamma$-ray production and can be used to test various models. 

\section{Acknowledgements}

The {\Fermi}-LAT Collaboration acknowledges generous ongoing support
from a number of agencies and institutes that have supported both the
development and the operation of the LAT as well as scientific data analysis.
These include the National Aeronautics and Space Administration and the
Department of Energy in the United States, the Commissariat \`a l'Energie Atomique
and the Centre National de la Recherche Scientifique / Institut National de Physique
Nucl\'eaire et de Physique des Particules in France, the Agenzia Spaziale Italiana
and the Istituto Nazionale di Fisica Nucleare in Italy, the Ministry of Education,
Culture, Sports, Science and Technology (MEXT), High Energy Accelerator Research
Organization (KEK) and Japan Aerospace Exploration Agency (JAXA) in Japan, and
the K.~A.~Wallenberg Foundation, the Swedish Research Council and the
Swedish National Space Board in Sweden.

Additional support for science analysis during the operations phase is gratefully
acknowledged from the Istituto Nazionale di Astrofisica in Italy and the Centre National d'\'Etudes Spatiales in France.

This paper is based on observations obtained with the Samuel Oschin Telescope as part of the Palomar Transient Factory project, a scientific collaboration between the California Institute of Technology, Columbia University, Las Cumbres Observatory, the Lawrence Berkeley National Laboratory, the National Energy Research Scientific Computing Center,                                                                                                                                            
the University of Oxford, and the Weizmann Institute of Science. Some of the data presented herein were obtained at the W. M. Keck Observatory, which is operated as a scientific partnership among the California Institute of Technology, the University of California, and NASA; the Observatory was made possible by the generous financial support of the W. M. Keck Foundation.  We are grateful for excellent staff assistance at Palomar, Lick, and Keck Observatories.  E.O.O. is incumbent of the Arye Dissentshik career development chair and is grateful to support by grants from the Willner Family Leadership Institute Ilan Gluzman (Secaucus NJ), Israeli Ministry of Science, Israel Science Foundation, Minerva and the I-CORE Program of the Planning and Budgeting Committee and The Israel Science Foundation. A.~G.-Y. is supported by the EU/FP7 via ERC grant No. 307260, the Quantum Universe I-Core program by the Israeli Committee for planning and budgeting and the ISF, Minerva and ISF grants, WIS-UK ``making connections", and Kimmel and ARCHES awards. 
\appendix
\section{SN Catalog}
\label{sec:AppendixA}
The following table contains all SNe included in this analysis. The column definition is similar to Table~\ref{tab:closeSNshort}.
\begin{center}
{\footnotesize\begin{longtable}{ l c c c c c c }
\hline
\hline
Name & RA ($^\circ$)$^{\star}$ & Dec ($^\circ$)$^{\star}$ & Date  & z & m & TS (p-value) \\
\hline
\hline
SN2008gm & 348.55 & -2.78 & 2008-10-22$^\dagger$  & 0.012 & 17.00$^\ddagger$  & 3.2 (0.169) \\
\hline
CSS081201\_103354-032125 & 158.47 & -3.36 & 2008-12-01$^\dagger$  & 0.060 & 18.30$^\ddagger$  & 0.0 (0.572) \\
\hline
CSS080701\_234413+075224 & \multirow{2}{*}{356.05} & \multirow{2}{*}{7.87} & \multirow{2}{*}{2008-12-30} & \multirow{2}{*}{0.069} & \multirow{2}{*}{18.50} & \multirow{2}{*}{0.0 (0.572)} \\
SN2008ja & & & & & & \\
\hline
SN2008ip & 194.46 & 36.38 & 2008-12-31$^\dagger$  & 0.015 & 15.70$^\ddagger$  & 0.0 (0.572) \\
\hline
SN2009au & 194.94 & -29.60 & 2009-03-11$^\dagger$  & 0.009 & 16.40$^\ddagger$  & 0.0 (0.572) \\
\hline
CSS080928\_160837+041626 & \multirow{2}{*}{242.16} & \multirow{2}{*}{4.27} & \multirow{2}{*}{2009-03-21} & \multirow{2}{*}{0.041} & \multirow{2}{*}{17.60} & \multirow{2}{*}{0.3 (0.458)} \\
SN2008iy & & & & & & \\
\hline
SN2009cw & 226.26 & 48.67 & 2009-03-28$^\dagger$  & 0.150 & 20.30$^\ddagger$  & 0.0 (0.572) \\
\hline
SN2009eo & 224.53 & 2.43 & 2009-05-13$^\dagger$  & 0.044 & 18.10$^\ddagger$  & 0.0 (0.572) \\
\hline
SN2009fs & 274.80 & 42.81 & 2009-06-01$^\dagger$  & 0.054 & 17.00$^\ddagger$  & 3.5 (0.154) \\
\hline
PTF09ij & 218.06 & 54.86 & 2009-06-03 & 0.124 & 20.30 & 0.0 (0.572) \\
\hline
PTF09ge & 224.26 & 49.61 & 2009-06-06 & 0.064 & 17.90 & 3.3 (0.165) \\
\hline
PTF09tm & 206.73 & 61.55 & 2009-06-25 & 0.034 & 16.80 & 0.0 (0.572) \\
\hline
PTF09uj & 215.05 & 53.56 & 2009-06-26 & 0.066 & 18.20 & 0.0 (0.572) \\
\hline
PTF09uy & 190.98 & 74.69 & 2009-07-03 & 0.313 & 19.40 & 0.0 (0.572) \\
\hline
PTF09bcl & 271.61 & 17.86 & 2009-07-19$^\dagger$  & 0.062 & 20.87$^\ddagger$  & 0.0 (0.572) \\
\hline
PTF10ujc & 353.63 & 22.35 & 2009-08-05 & 0.032 & 16.20 & 0.0 (0.572) \\
\hline
PTF09drs & 226.63 & 60.59 & 2009-08-15 & 0.045 & 18.50 & 0.0 (0.561) \\
\hline
CSS090925\_001259+144121 & 3.25 & 14.69 & 2009-09-25$^\dagger$  & 0.090 & 18.80$^\ddagger$  & 0.0 (0.568) \\
\hline
SN2009ma & 127.24 & 0.59 & 2009-10-17$^\dagger$  & 0.089 & 18.20$^\ddagger$  & 0.0 (0.572) \\
\hline
CSS091018\_091109+195945 & \multirow{2}{*}{137.79} & \multirow{2}{*}{20.00} & \multirow{2}{*}{2009-10-18}$^\dagger$  & \multirow{2}{*}{0.150} & \multirow{2}{*}{19.00}$^\ddagger$  & \multirow{2}{*}{0.0 (0.572)} \\
SN2009mb & & & & & & \\
\hline
SN2009kn & 122.43 & -17.75 & 2009-10-26$^\dagger$  & 0.016 & 16.60$^\ddagger$  & 0.0 (0.572) \\
\hline
SN2009kr & 78.01 & -15.70 & 2009-11-06$^\dagger$  & 0.006 & 16.00$^\ddagger$  & 4.7 (0.104) \\
\hline
SN2009nm & 151.35 & 51.28 & 2009-11-20$^\dagger$  & 0.210 & 18.80$^\ddagger$  & 0.0 (0.572) \\
\hline
CSS091217\_110637+341952 & \multirow{2}{*}{166.65} & \multirow{2}{*}{34.33} & \multirow{2}{*}{2009-12-17}$^\dagger$  & \multirow{2}{*}{?} & \multirow{2}{*}{18.70}$^\ddagger$  & \multirow{2}{*}{0.0 (0.572)} \\
SN2009nj & & & & & & \\
\hline
CSS091218\_104011+223735 & \multirow{2}{*}{160.05} & \multirow{2}{*}{22.63} & \multirow{2}{*}{2009-12-18}$^\dagger$  & \multirow{2}{*}{0.140} & \multirow{2}{*}{19.40}$^\ddagger$  & \multirow{2}{*}{0.0 (0.572)} \\
SN2009nw & & & & & & \\
\hline
PTF10dk & 77.09 & 0.21 & 2009-12-18$^\dagger$  & 0.074 & 20.14$^\ddagger$  & 0.0 (0.572) \\
\hline
PTF10u & 152.49 & 46.01 & 2010-01-05 & 0.150 & 19.80 & 0.0 (0.572) \\
\hline
PTF11ner & 125.58 & 72.83 & 2010-01-11$^\dagger$  & 0.117 & 20.94$^\ddagger$  & 0.0 (0.572) \\
\hline
PTF10ts & \multirow{2}{*}{188.49} & \multirow{2}{*}{13.92} & \multirow{2}{*}{2010-01-12} & \multirow{2}{*}{0.046} & \multirow{2}{*}{17.66} & \multirow{2}{*}{7.9 (0.033)} \\
SN2009nn & & & & & & \\
\hline
CSS100113\_032138+263650 & \multirow{2}{*}{50.41} & \multirow{2}{*}{26.61} & \multirow{2}{*}{2010-01-13}$^\dagger$  & \multirow{2}{*}{0.060} & \multirow{2}{*}{18.80}$^\ddagger$  & \multirow{2}{*}{0.1 (0.517)} \\
SN2010M & & & & & & \\
\hline
PTF10cwl & 189.09 & 7.79 & 2010-03-13$^\dagger$  & 0.085 & 19.00$^\ddagger$  & 0.0 (0.572) \\
\hline
SN2010al & 123.57 & 18.44 & 2010-03-13$^\dagger$  & 0.017 & 17.80$^\ddagger$  & 9.7 (0.023) \\
\hline
PTF10cwx & 188.32 & -0.05 & 2010-03-19 & 0.073 & 18.50 & 2.3 (0.228) \\
\hline
PTF10fei & 227.07 & 53.59 & 2010-04-04 & 0.090 & 18.55 & 0.0 (0.572) \\
\hline
PTF10fel & 246.88 & 51.36 & 2010-04-04 & 0.234 & 19.70 & 11.1 (0.016) \\
\hline
PTF10ewc & 210.50 & 33.84 & 2010-04-15 & 0.055 & 18.40 & 0.3 (0.476) \\
\hline
PTF10fou & 208.94 & 29.88 & 2010-04-17 & 0.043 & 20.00 & 0.2 (0.489) \\
\hline
PTF10flx & 251.74 & 64.45 & 2010-04-17 & 0.067 & 18.80 & 11.6 (0.015) \\
\hline
SN2010bt & 192.08 & -34.95 & 2010-04-17$^\dagger$  & 0.016 & 15.80$^\ddagger$  & 14.4 (0.0065) \\
\hline
PTF10fjh & \multirow{2}{*}{251.73} & \multirow{2}{*}{34.16} & \multirow{2}{*}{2010-04-25} & \multirow{2}{*}{0.032} & \multirow{2}{*}{17.20} & \multirow{2}{*}{0.0 (0.572)} \\
SN2010bq & & & & & & \\
\hline
PTF10gvd & 253.26 & 67.00 & 2010-05-02 & 0.070 & 19.20 & 2.8 (0.196) \\
\hline
PTF10hcr & 183.00 & 38.53 & 2010-05-06$^\dagger$  & 0.037 & 20.06$^\ddagger$  & 1.0 (0.359) \\
\hline
PTF10hbf & 193.19 & -6.92 & 2010-05-07 & 0.042 & 18.80 & 0.6 (0.407) \\
\hline
PTF10hif & 257.45 & 27.26 & 2010-05-12 & 0.141 & 18.00 & 0.0 (0.572) \\
\hline
PTF10gvf & 168.44 & 53.63 & 2010-05-14 & 0.080 & 19.00 & 0.0 (0.572) \\
\hline
PTF10hSN & 244.40 & 5.04 & 2010-06-01 & 0.164 & 19.00 & 0.0 (0.572) \\
\hline
PTF10jop & 322.38 & 2.88 & 2010-06-11 & 0.089 & 18.60 & 0.0 (0.572) \\
\hline
PTF10ngx & 186.80 & 15.98 & 2010-07-03 & 0.067 & 19.40 & 0.0 (0.572) \\
\hline
PTF10ndr & 224.95 & 65.00 & 2010-07-26 & 0.075 & 19.60 & 0.0 (0.572) \\
\hline
PTF10qaf & 353.93 & 10.78 & 2010-07-31 & 0.284 & 19.00 & 7.5 (0.036) \\
\hline
SN2010hd & 340.47 & -46.10 & 2010-08-07$^\dagger$  & 0.033 & 17.60$^\ddagger$  & 8.8 (0.028) \\
\hline
PS1-1000789 & 310.69 & 15.51 & 2010-08-15$^\dagger$  & 0.200 & 17.30$^\ddagger$  & 0.0 (0.572) \\
\hline
PTF10oug & 260.19 & 29.07 & 2010-08-20 & 0.150 & 19.20 & 0.0 (0.572) \\
\hline
PTF10scc & 352.04 & 28.64 & 2010-08-20 & 0.242 & 18.90 & 0.0 (0.572) \\
\hline
PTF10qwu & 252.79 & 28.30 & 2010-08-20 & 0.226 & 19.40 & 0.1 (0.541) \\
\hline
PTF10tjr & 220.38 & 23.01 & 2010-08-23$^\dagger$  & 0.078 & 17.73$^\ddagger$  & 0.0 (0.572) \\
\hline
PTF10tpz & 329.63 & -15.55 & 2010-08-28$^\dagger$  & 0.040 & 17.06$^\ddagger$  & 0.0 (0.572) \\
\hline
PTF10tel & \multirow{2}{*}{260.38} & \multirow{2}{*}{48.13} & \multirow{2}{*}{2010-09-04} & \multirow{2}{*}{0.035} & \multirow{2}{*}{17.50} & \multirow{2}{*}{0.0 (0.572)} \\
SN2010mc & & & & & & \\
\hline
PTF10ttp & 341.92 & -10.04 & 2010-09-09 & 0.179 & 19.50 & 0.0 (0.572) \\
\hline
CSS100910\_001539+271250 & 3.91 & 27.21 & 2010-09-10$^\dagger$  & 0.024 & 18.10$^\ddagger$  & 0.0 (0.572) \\
\hline
PTF10viv & \multirow{2}{*}{331.11} & \multirow{2}{*}{-7.98} & \multirow{2}{*}{2010-09-12}$^\dagger$  & \multirow{2}{*}{0.060} & \multirow{2}{*}{20.13}$^\ddagger$  & \multirow{2}{*}{0.0 (0.572)} \\
SN2010jg & & & & & & \\
\hline
PTF10uls & 20.34 & 4.89 & 2010-09-19 & 0.044 & 18.60 & 1.4 (0.322) \\
\hline
PTF10xzs & 120.60 & 67.42 & 2010-09-22$^\dagger$  & 0.036 & 19.33$^\ddagger$  & 8.2 (0.031) \\
\hline
PTF10wop & 327.65 & -6.77 & 2010-09-23$^\dagger$  & 0.090 & 19.55$^\ddagger$  & 0.0 (0.572) \\
\hline
PTF10xif & 48.11 & -9.81 & 2010-09-27$^\dagger$  & 0.029 & 18.42$^\ddagger$  & 0.0 (0.572) \\
\hline
PTF10vag & 326.83 & 18.13 & 2010-09-29 & 0.052 & 18.50 & 0.0 (0.572) \\
\hline
PTF10xgo & 328.99 & 1.32 & 2010-10-03$^\dagger$  & 0.034 & 19.25$^\ddagger$  & 2.8 (0.193) \\
\hline
CSS121009\_025917-141610 & 44.82 & -14.27 & 2010-10-09$^\dagger$  & 0.080 & 19.20$^\ddagger$  & 0.0 (0.572) \\
\hline
PTF10tyd & 257.33 & 27.82 & 2010-10-09 & 0.063 & 19.00 & 0.0 (0.572) \\
\hline
PTF12kph & 24.82 & -7.56 & 2010-10-11$^\dagger$  & 0.059 & 18.84$^\ddagger$  & 6.1 (0.063) \\
\hline
PTF10uiz & 258.63 & 21.43 & 2010-10-19 & 0.114 & 18.40 & 0.0 (0.572) \\
\hline
PTF10wmk & 132.04 & 55.83 & 2010-10-29 & 0.137 & 19.51 & 0.0 (0.572) \\
\hline
PTF10yzt & 2.96 & 26.69 & 2010-10-29$^\dagger$  & 0.076 & 18.58$^\ddagger$  & 0.0 (0.572) \\
\hline
CSS101030\_230944+054156 & \multirow{2}{*}{347.43} & \multirow{2}{*}{5.70} & \multirow{2}{*}{2010-10-30}$^\dagger$  & \multirow{2}{*}{0.042} & \multirow{2}{*}{16.50}$^\ddagger$  & \multirow{2}{*}{0.0 (0.572)} \\
SN2010jy & & & & & & \\
\hline
PTF10aaes & 31.79 & 16.21 & 2010-10-30$^\dagger$  & 0.037 & 19.50$^\ddagger$  & 0.0 (0.572) \\
\hline
SN2010jk & 18.15 & 15.47 & 2010-10-31$^\dagger$  & 0.280 & 20.20$^\ddagger$  & 3.6 (0.153) \\
\hline
PTF10acfd & 147.91 & 1.52 & 2010-11-03$^\dagger$  & 0.192 & 20.34$^\ddagger$  & 0.0 (0.572) \\
\hline
SN2010lx & 71.19 & -22.21 & 2010-11-03$^\dagger$  & 0.100 & 18.70$^\ddagger$  & 0.0 (0.572) \\
\hline
SN2010js & 124.21 & 60.50 & 2010-11-07$^\dagger$  & 0.039 & 18.10$^\ddagger$  & 1.9 (0.262) \\
\hline
PTF10yyc & 69.82 & -0.35 & 2010-11-08 & 0.214 & 17.66 & 0.9 (0.367) \\
\hline
PTF10weh & 261.71 & 58.85 & 2010-11-08 & 0.138 & 18.30 & 6.6 (0.048) \\
\hline
CSS101110\_082047+355337 & \multirow{2}{*}{125.20} & \multirow{2}{*}{35.89} & \multirow{2}{*}{2010-11-10}$^\dagger$  & \multirow{2}{*}{0.075} & \multirow{2}{*}{18.20}$^\ddagger$  & \multirow{2}{*}{6.2 (0.059)} \\
2010kb & & & & & & \\
\hline
PTF10aazn & \multirow{2}{*}{31.72} & \multirow{2}{*}{44.57} & \multirow{2}{*}{2010-11-13}$^\dagger$  & \multirow{2}{*}{0.016} & \multirow{2}{*}{16.52}$^\ddagger$  & \multirow{2}{*}{5.4 (0.079)} \\
SN2010jj & & & & & & \\
\hline
PTF10aaxf & \multirow{2}{*}{145.72} & \multirow{2}{*}{9.50} & \multirow{2}{*}{2010-11-18} & \multirow{2}{*}{0.011} & \multirow{2}{*}{13.20} & \multirow{2}{*}{7.1 (0.039)} \\
SN2010jl & & & & & & \\
\hline
PTF10abcl & 348.90 & 22.81 & 2010-11-19$^\dagger$  & 0.061 & 18.95$^\ddagger$  & 3.0 (0.181) \\
\hline
PTF10aaxi & \multirow{2}{*}{94.13} & \multirow{2}{*}{-21.41} & \multirow{2}{*}{2010-11-23} & \multirow{2}{*}{0.010} & \multirow{2}{*}{18.00} & \multirow{2}{*}{0.0 (0.572)} \\
SN2010jp & & & & & & \\
\hline
PTF10yni & 2.71 & 14.18 & 2010-11-28 & 0.169 & 18.90 & 7.2 (0.039) \\
\hline
PTF10abui & 93.08 & -22.77 & 2010-12-08 & 0.052 & 18.60 & 0.0 (0.572) \\
\hline
PTF10abyy & 79.17 & 6.80 & 2010-12-08$^\dagger$  & 0.030 & 18.66$^\ddagger$  & 0.2 (0.503) \\
\hline
PTF10achk & 46.49 & -10.52 & 2010-12-28 & 0.033 & 16.90 & 0.0 (0.572) \\
\hline
SN2011A & 195.25 & -14.53 & 2011-01-02$^\dagger$  & 0.009 & 16.90$^\ddagger$  & 0.0 (0.572) \\
\hline
SN2011P & 36.44 & 16.22 & 2011-01-05$^\dagger$  & 0.080 & 18.60$^\ddagger$  & 10.8 (0.016) \\
\hline
SN2011af & 36.48 & 10.39 & 2011-01-11$^\dagger$  & 0.064 & 16.70$^\ddagger$  & 0.0 (0.572) \\
\hline
SN2011S & 138.48 & -17.01 & 2011-01-14$^\dagger$  & 0.060 & 17.60$^\ddagger$  & 0.0 (0.572) \\
\hline
PTF10acsq & 120.39 & 46.76 & 2011-01-27 & 0.173 & 19.00 & 0.0 (0.572) \\
\hline
SN2011ap & 272.62 & 31.01 & 2011-02-21$^\dagger$  & 0.024 & 18.30$^\ddagger$  & 0.0 (0.572) \\
\hline
SN2011an & 119.85 & 16.42 & 2011-03-01$^\dagger$  & 0.016 & 18.40$^\ddagger$  & 0.7 (0.398) \\
\hline
SN2011cc & 248.46 & 39.26 & 2011-03-17$^\dagger$  & 0.032 & 17.70$^\ddagger$  & 0.0 (0.572) \\
\hline
PS1-11xn & 221.91 & 51.68 & 2011-04-26$^\dagger$  & 0.040 & 18.60$^\ddagger$  & 0.0 (0.572) \\
\hline
SN2011cp & 118.14 & 21.89 & 2011-04-26$^\dagger$  & 0.390 & 19.50$^\ddagger$  & 5.4 (0.081) \\
\hline
CSS110501\_094825+204333 & 147.10 & 20.73 & 2011-05-01$^\dagger$  & 0.040 & 18.40$^\ddagger$  & 0.0 (0.572) \\
\hline
PTF11csc & 224.68 & 36.60 & 2011-05-02 & 0.117 & 20.60 & 0.0 (0.572) \\
\hline
PTF11dsb & 244.65 & 32.70 & 2011-05-15 & 0.190 & 20.10 & 0.0 (0.572) \\
\hline
SN2011eu & 212.31 & -1.18 & 2011-06-06$^\dagger$  & 0.110 & 18.50$^\ddagger$  & 0.0 (0.572) \\
\hline
PTF11fuu & 325.12 & 6.33 & 2011-06-09 & 0.097 & 18.50 & 0.0 (0.572) \\
\hline
PTF11fss & 323.47 & 1.84 & 2011-06-11$^\dagger$  & 0.125 & 19.42$^\ddagger$  & 0.0 (0.572) \\
\hline
CSS110623\_131919-045106 & 199.83 & -4.85 & 2011-06-23$^\dagger$  & 0.070 & 18.40$^\ddagger$  & 0.0 (0.572) \\
\hline
PTF11gtr & 258.01 & 23.38 & 2011-06-25$^\dagger$  & 0.029 & 20.94$^\ddagger$  & 0.0 (0.572) \\
\hline
PTF11hzx & 327.67 & 18.11 & 2011-07-24 & 0.229 & 18.90 & 0.0 (0.572) \\
\hline
PTF11iqb & 8.52 & -9.70 & 2011-08-06 & 0.013 & 15.20 & 0.3 (0.469) \\
\hline
PTF11fzz & 167.69 & 54.11 & 2011-08-15 & 0.082 & 17.40 & 0.3 (0.479) \\
\hline
SN2011fh & 194.06 & -29.50 & 2011-08-24$^\dagger$  & 0.008 & 14.50$^\ddagger$  & 1.9 (0.262) \\
\hline
PTF11pab & 44.63 & 6.31 & 2011-08-30$^\dagger$  & 0.022 & 21.08$^\ddagger$  & 6.1 (0.063) \\
\hline
SN2011fx & 4.50 & 24.56 & 2011-08-30$^\dagger$  & 0.019 & 17.60$^\ddagger$  & 0.5 (0.428) \\
\hline
SN2011fr & 22.44 & 18.89 & 2011-09-01$^\dagger$  & 0.060 & 18.80$^\ddagger$  & 0.0 (0.572) \\
\hline
PTF11mpg & 334.40 & 0.61 & 2011-09-19 & 0.093 & 19.18 & 0.0 (0.572) \\
\hline
PTF11oey & 352.73 & 23.18 & 2011-09-21$^\dagger$  & 0.061 & 20.17$^\ddagger$  & 0.0 (0.572) \\
\hline
PTF11mtq & 270.08 & 28.70 & 2011-09-22$^\dagger$  & 0.073 & 19.35$^\ddagger$  & 1.6 (0.302) \\
\hline
PTF11msk & 325.91 & -1.69 & 2011-10-04 & 0.070 & 19.10 & 2.2 (0.238) \\
\hline
PTF11pdt & 44.63 & 6.31 & 2011-10-19$^\dagger$  & 0.022 & 20.00$^\ddagger$  & 8.9 (0.028) \\
\hline
PSNJ10081059+5150570 & \multirow{2}{*}{152.04} & \multirow{2}{*}{51.85} & \multirow{2}{*}{2011-10-29} & \multirow{2}{*}{0.004} & \multirow{2}{*}{14.50} & \multirow{2}{*}{0.0 (0.572)} \\
SN2011ht & & & & & & \\
\hline
PTF11qnf & 86.23 & 69.15 & 2011-11-01$^\dagger$  & 0.014 & 19.80$^\ddagger$  & 1.4 (0.320) \\
\hline
SN2011ib & 176.16 & 35.97 & 2011-11-15$^\dagger$  & 0.037 & 16.80$^\ddagger$  & 0.0 (0.572) \\
\hline
SN2011hw & 336.56 & 34.22 & 2011-11-18$^\dagger$  & 0.023 & 15.70$^\ddagger$  & 0.0 (0.572) \\
\hline
SN2011jb & 174.27 & 15.47 & 2011-11-28$^\dagger$  & 0.084 & 17.80$^\ddagger$  & 0.0 (0.572) \\
\hline
SN2011iw & 353.70 & 24.75 & 2011-11-29$^\dagger$  & 0.023 & 16.90$^\ddagger$  & 0.0 (0.572) \\
\hline
PTF11qqj & 149.51 & 0.72 & 2011-12-11 & 0.093 & 19.00 & 0.0 (0.572) \\
\hline
PTF11oxu & \multirow{2}{*}{54.64} & \multirow{2}{*}{22.55} & \multirow{2}{*}{2011-12-13} & \multirow{2}{*}{0.088} & \multirow{2}{*}{18.70} & \multirow{2}{*}{4.7 (0.103)} \\
SN2011jc & & & & & & \\
\hline
PTF11rlv & 192.39 & -9.34 & 2011-12-21$^\dagger$  & 0.132 & 19.77$^\ddagger$  & 0.0 (0.572) \\
\hline
PTF11rfr & 25.57 & 29.27 & 2011-12-23 & 0.067 & 17.30 & 0.0 (0.572) \\
\hline
PTF12th & \multirow{2}{*}{72.62} & \multirow{2}{*}{-3.49} & \multirow{2}{*}{2012-01-05}$^\dagger$  & \multirow{2}{*}{0.084} & \multirow{2}{*}{19.08}$^\ddagger$  & \multirow{2}{*}{0.0 (0.572)} \\
SN2012Y & & & & & & \\
\hline
PTF12xv & 70.20 & 6.52 & 2012-01-18$^\dagger$  & 0.120 & 19.51$^\ddagger$  & 0.0 (0.572) \\
\hline
SN2012ab & 185.70 & 5.61 & 2012-01-31$^\dagger$  & 0.018 & 15.80$^\ddagger$  & 0.0 (0.572) \\
\hline
SN2012as & 231.29 & 37.96 & 2012-02-17$^\dagger$  & 0.029 & 17.90$^\ddagger$  & 0.0 (0.572) \\
\hline
SN2012al & 151.55 & 47.29 & 2012-02-24$^\dagger$  & 0.040 & 18.10$^\ddagger$  & 0.7 (0.398) \\
\hline
SN2012am & 163.51 & 46.03 & 2012-02-24$^\dagger$  & 0.042 & 17.60$^\ddagger$  & 0.0 (0.572) \\
\hline
LSQ12biu & 214.84 & -19.84 & 2012-03-21$^\dagger$  & 0.136 & 19.40$^\ddagger$  & 1.1 (0.352) \\
\hline
CSS120327\_110520-015205 & 166.33 & -1.87 & 2012-03-27$^\dagger$  & 0.090 & 17.80$^\ddagger$  & 0.0 (0.572) \\
\hline
CSS120330\_101639-064636 & \multirow{2}{*}{154.16} & \multirow{2}{*}{-6.78} & \multirow{2}{*}{2012-03-28}$^\dagger$  & \multirow{2}{*}{0.042} & \multirow{2}{*}{17.30}$^\ddagger$  & \multirow{2}{*}{0.0 (0.572)} \\
LSQ12by & & & & & & \\
\hline
PTF11mhr & 236.51 & 31.94 & 2012-03-28 & 0.054 & 17.30 & 0.4 (0.451) \\
\hline
LSQ12bqd & 197.91 & -16.40 & 2012-03-29$^\dagger$  & 0.041 & 19.30$^\ddagger$  & 0.0 (0.572) \\
\hline
SN2012bq & 154.16 & -6.78 & 2012-03-30$^\dagger$  & 0.042 & 17.60$^\ddagger$  & 0.0 (0.572) \\
\hline
PTF12cix & 191.29 & 35.94 & 2012-04-01 & 0.190 & 19.50 & 0.0 (0.572) \\
\hline
PTF12csy & 104.64 & 17.26 & 2012-04-07$^\dagger$  & 0.067 & 19.20$^\ddagger$  & 0.0 (0.572) \\
\hline
LSQ12btw\footnote{This source is of Type Ibn, while all other sources are of Type IIn.} & 152.62 & 5.54 & 2012-04-09$^\dagger$  & 0.057 & 19.10$^\ddagger$  & 0.0 (0.572) \\
\hline
PSNJ18410706-4147374 & \multirow{2}{*}{280.28} & \multirow{2}{*}{-41.79} & \multirow{2}{*}{2012-04-25}$^\dagger$  & \multirow{2}{*}{0.019} & \multirow{2}{*}{14.50}$^\ddagger$  & \multirow{2}{*}{0.0 (0.572)} \\
SN2012ca & & & & & & \\
\hline
PTF12cxj & 198.16 & 46.49 & 2012-04-28 & 0.035 & 18.70 & 0.0 (0.572) \\
\hline
\hline
\end{longtable}
}\end{center}

\section{Likelihood Profiles in Energy Bins}
\label{sec:LLHProfiles}

\begin{figure}[htbp]
\begin{center}
\includegraphics[scale=\twopic]{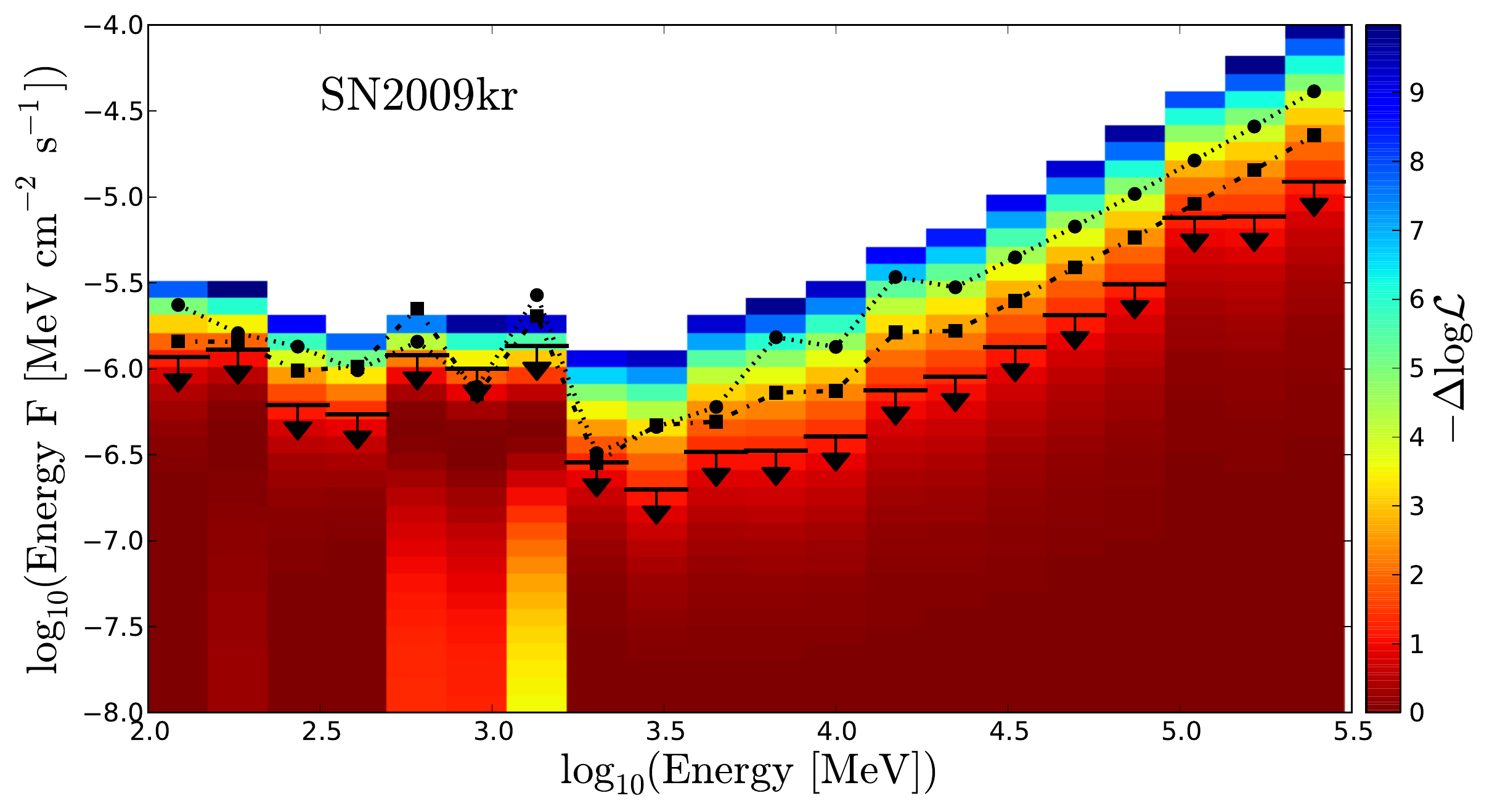}
\includegraphics[scale=\twopic]{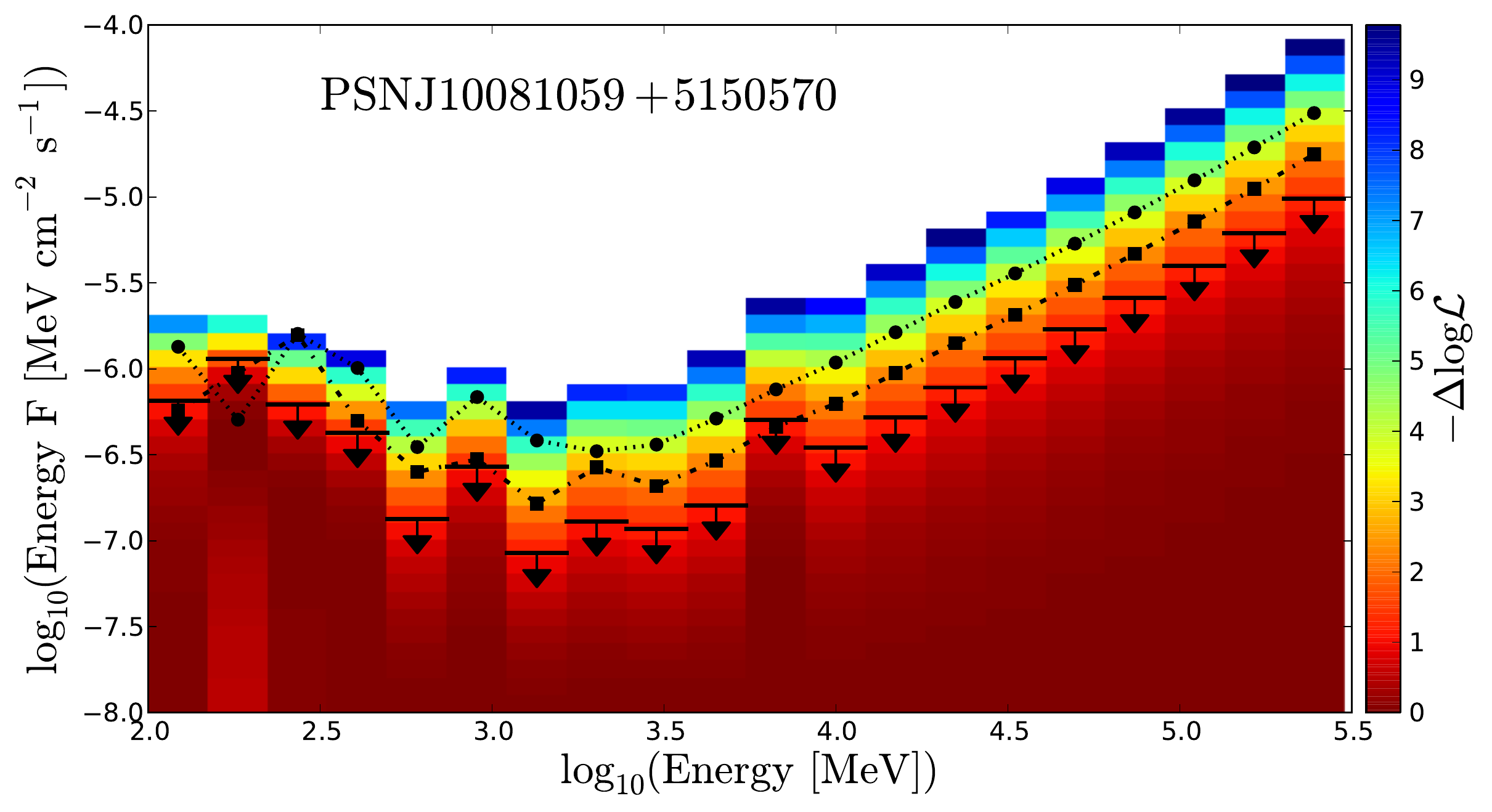}
\includegraphics[scale=\twopic]{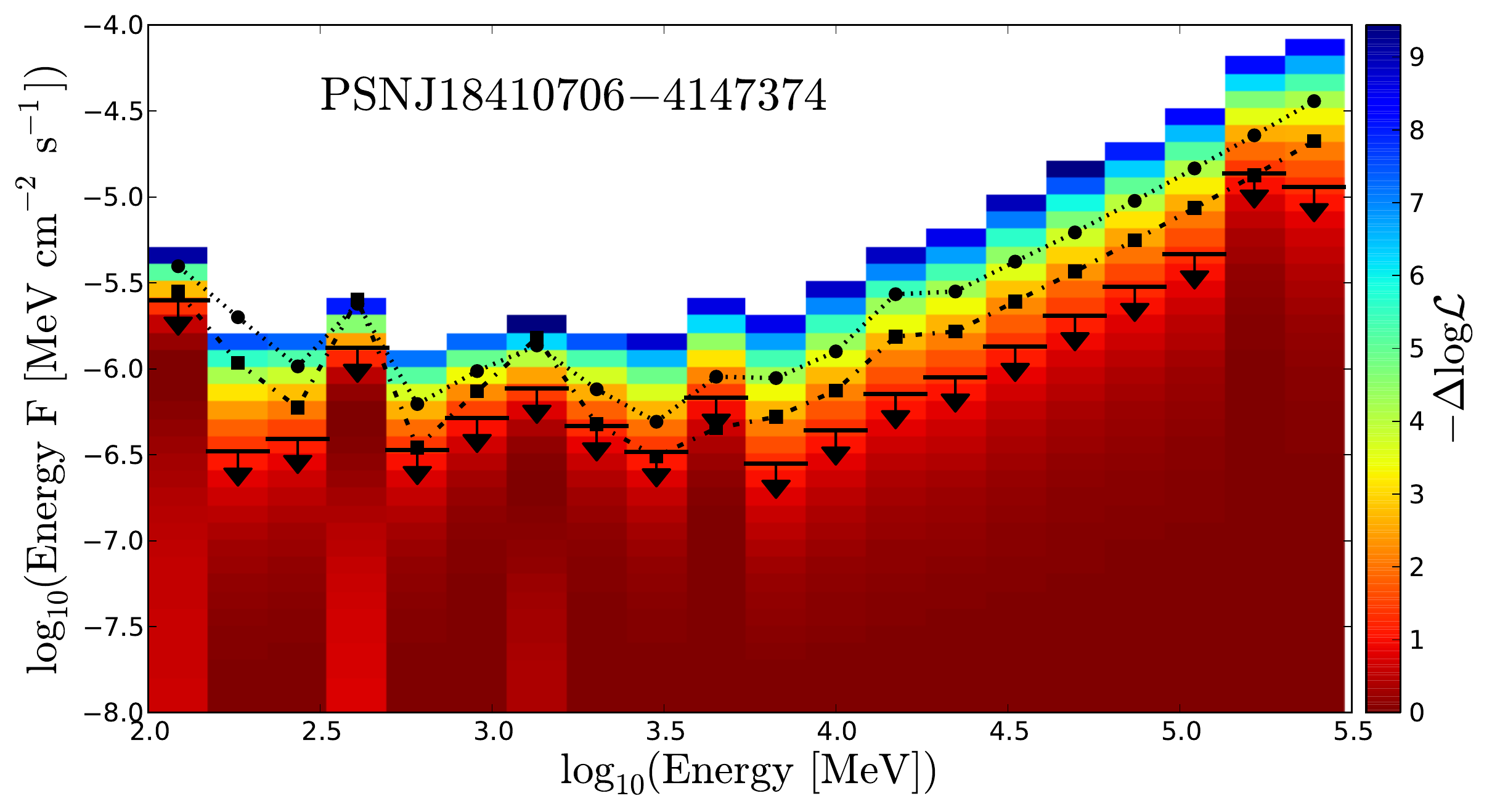}
\includegraphics[scale=\twopic]{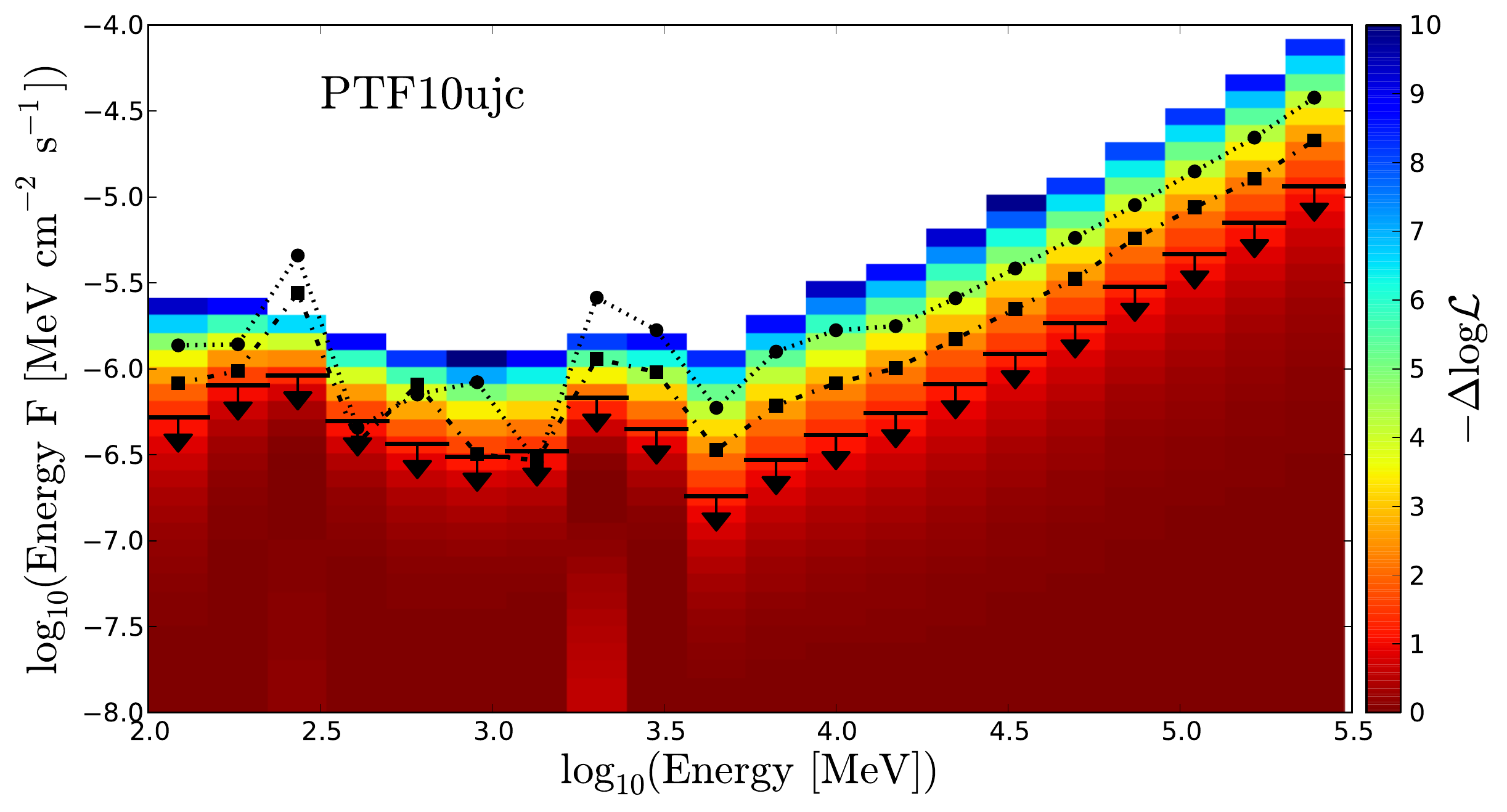}
\includegraphics[scale=\twopic]{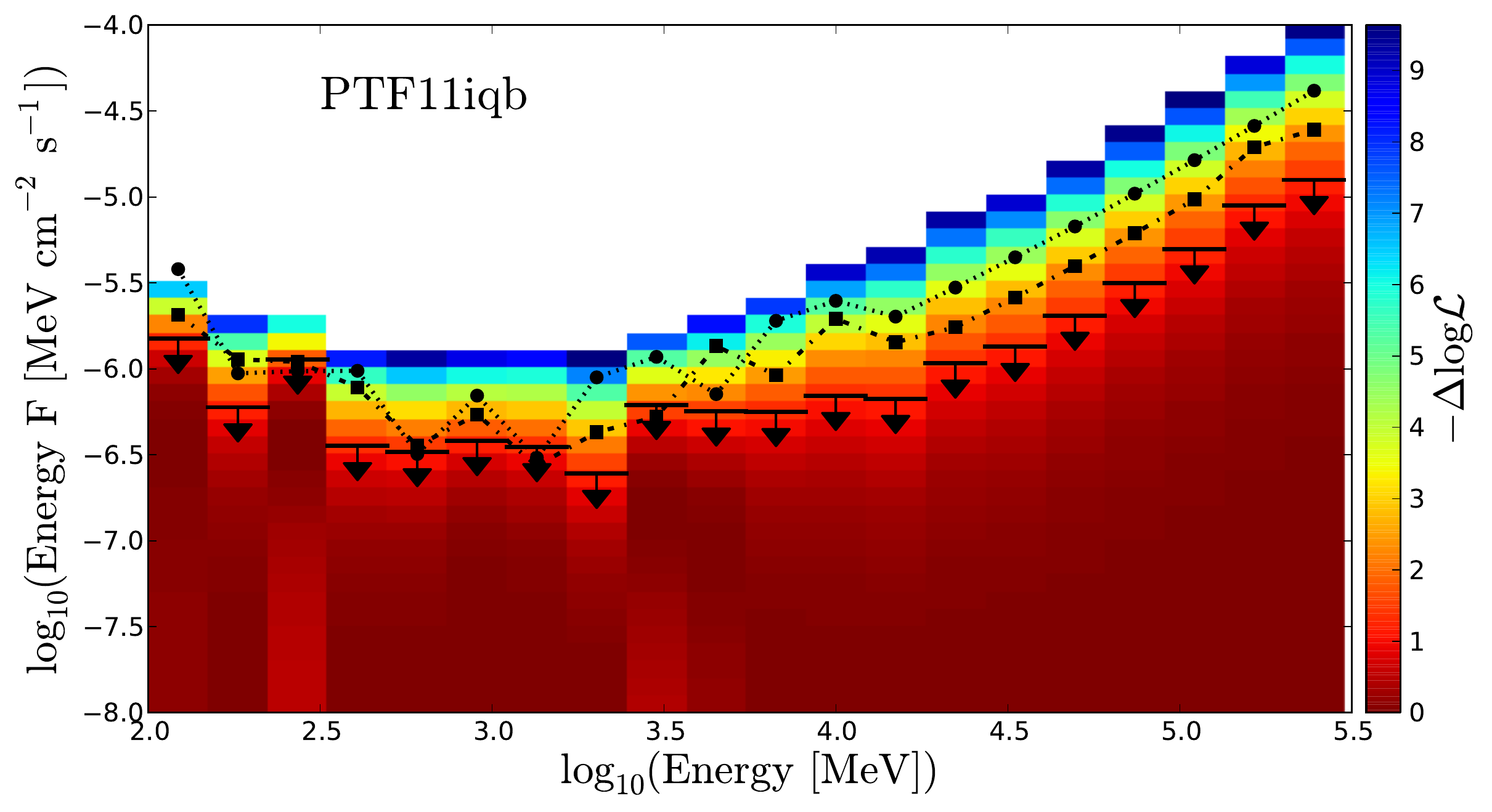}
\includegraphics[scale=\twopic]{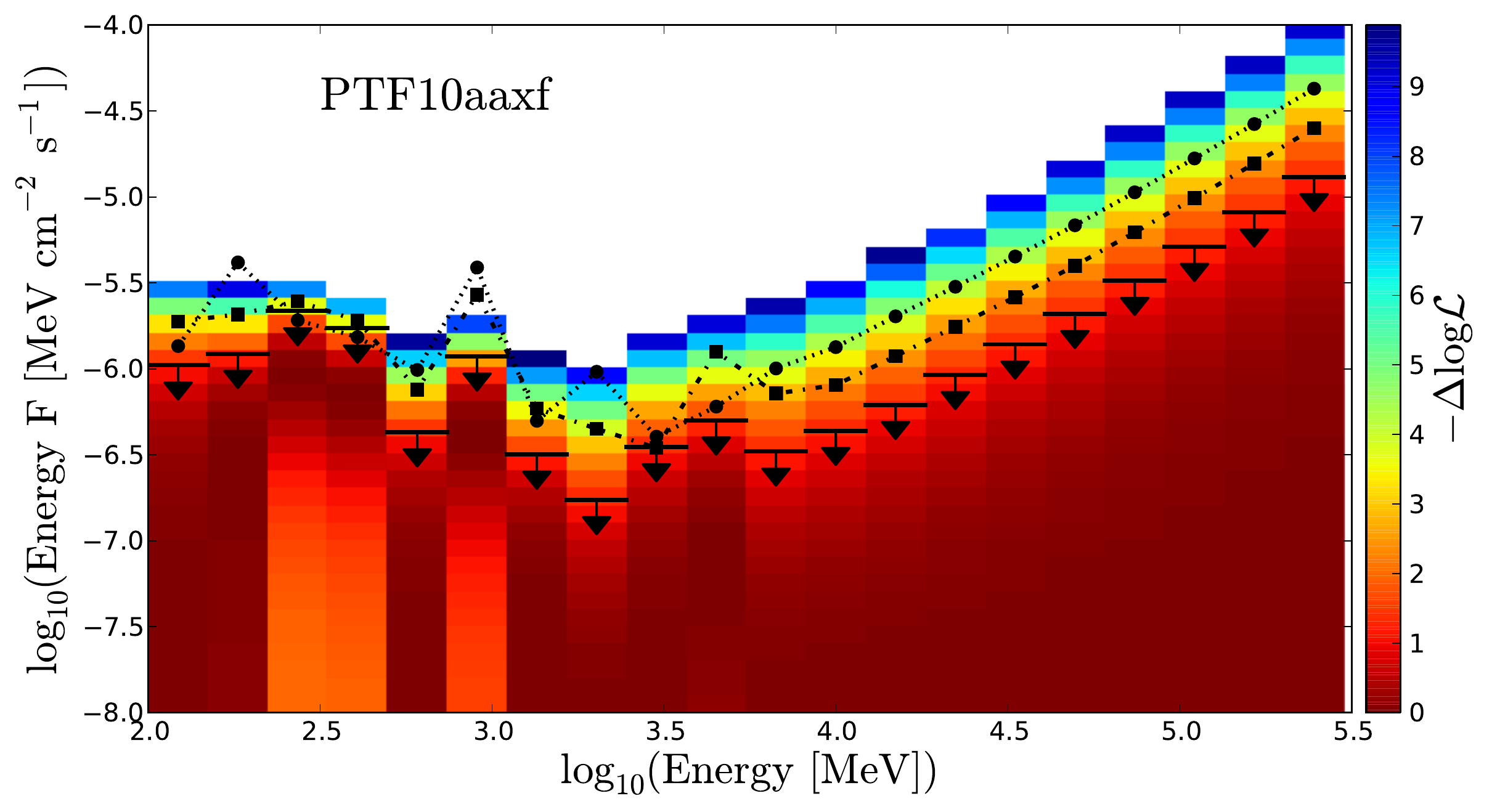}
\includegraphics[scale=\twopic]{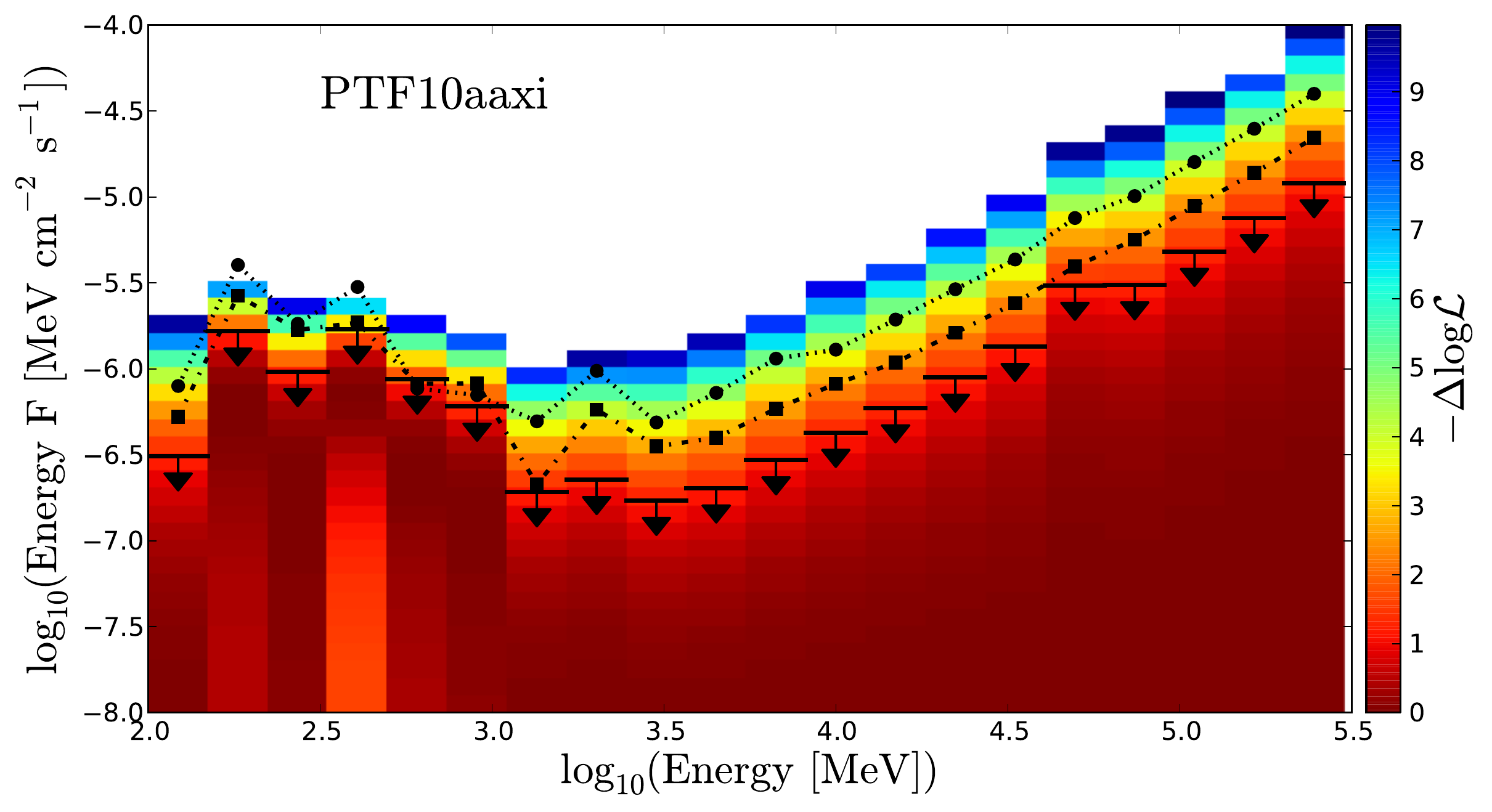}
\includegraphics[scale=\twopic]{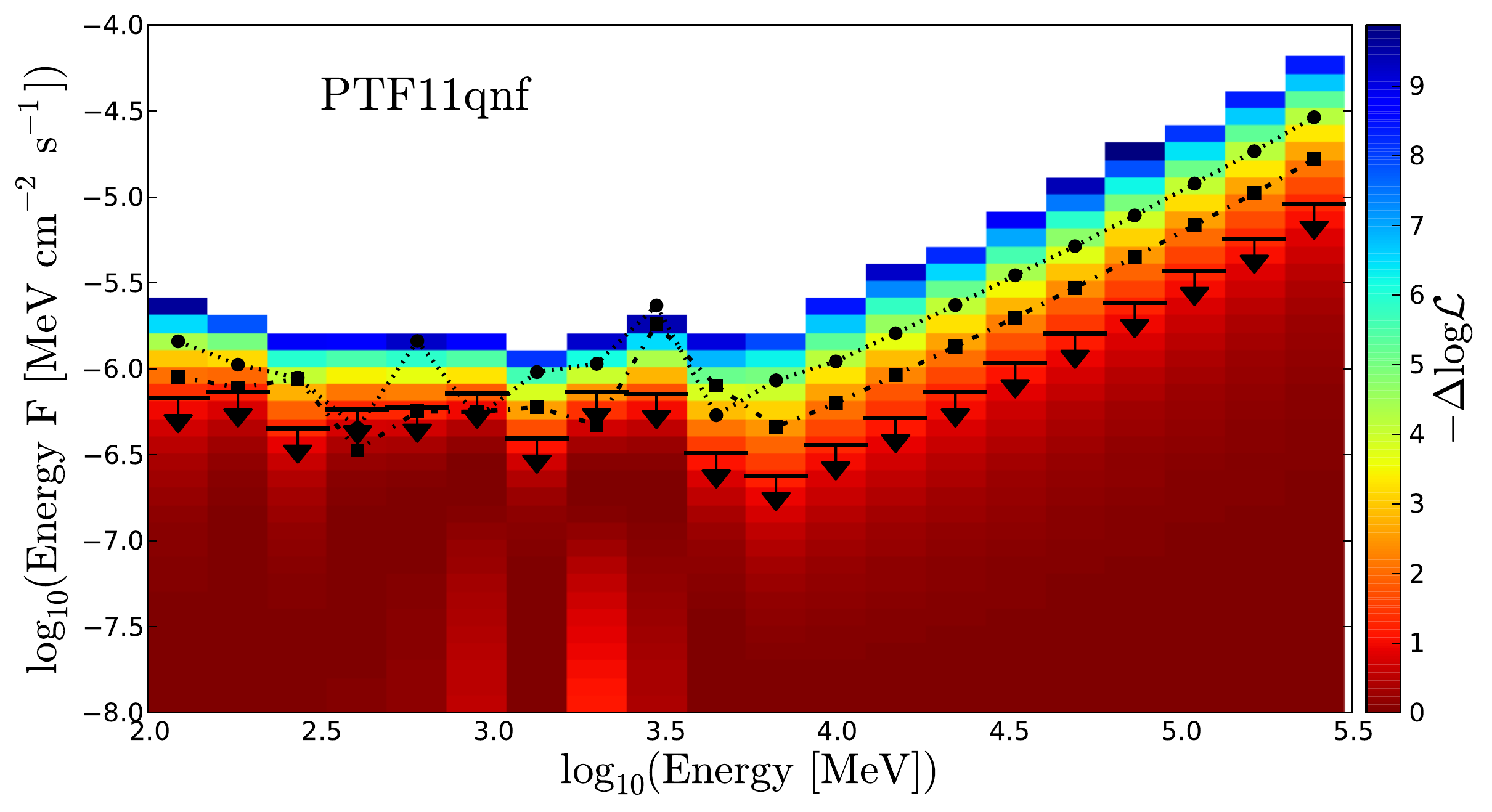}
\noindent
\caption{\small 
Similar to Fig.~\ref{fig:BinLLH}. Colors represent the likelihood profile for each energy for $\Delta T = 1$\,yr. The black arrows indicate the 95\% CL upper limits for $\Delta T = 1$\,yr, while the dotted-dashed and dotted lines represent the 95\% CL upper limit for $~\Delta T = 6$\,months and $\Delta T = 3$\,months respectively.}
\label{fig:BinLLHA}
\end{center}
\vspace{1mm}
\end{figure}

\begin{figure}[htbp]
\begin{center}
\includegraphics[scale=\twopic]{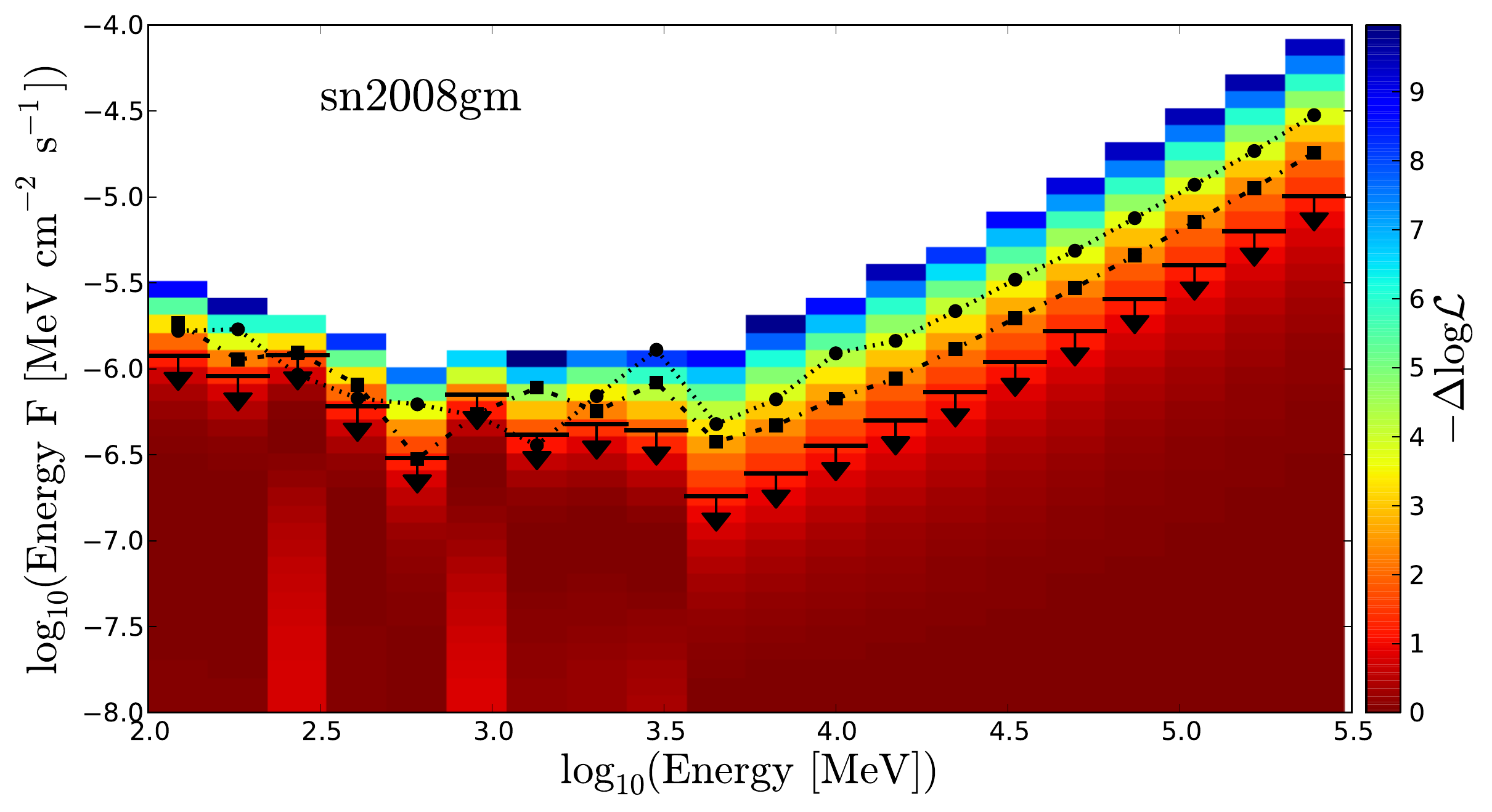}
\includegraphics[scale=\twopic]{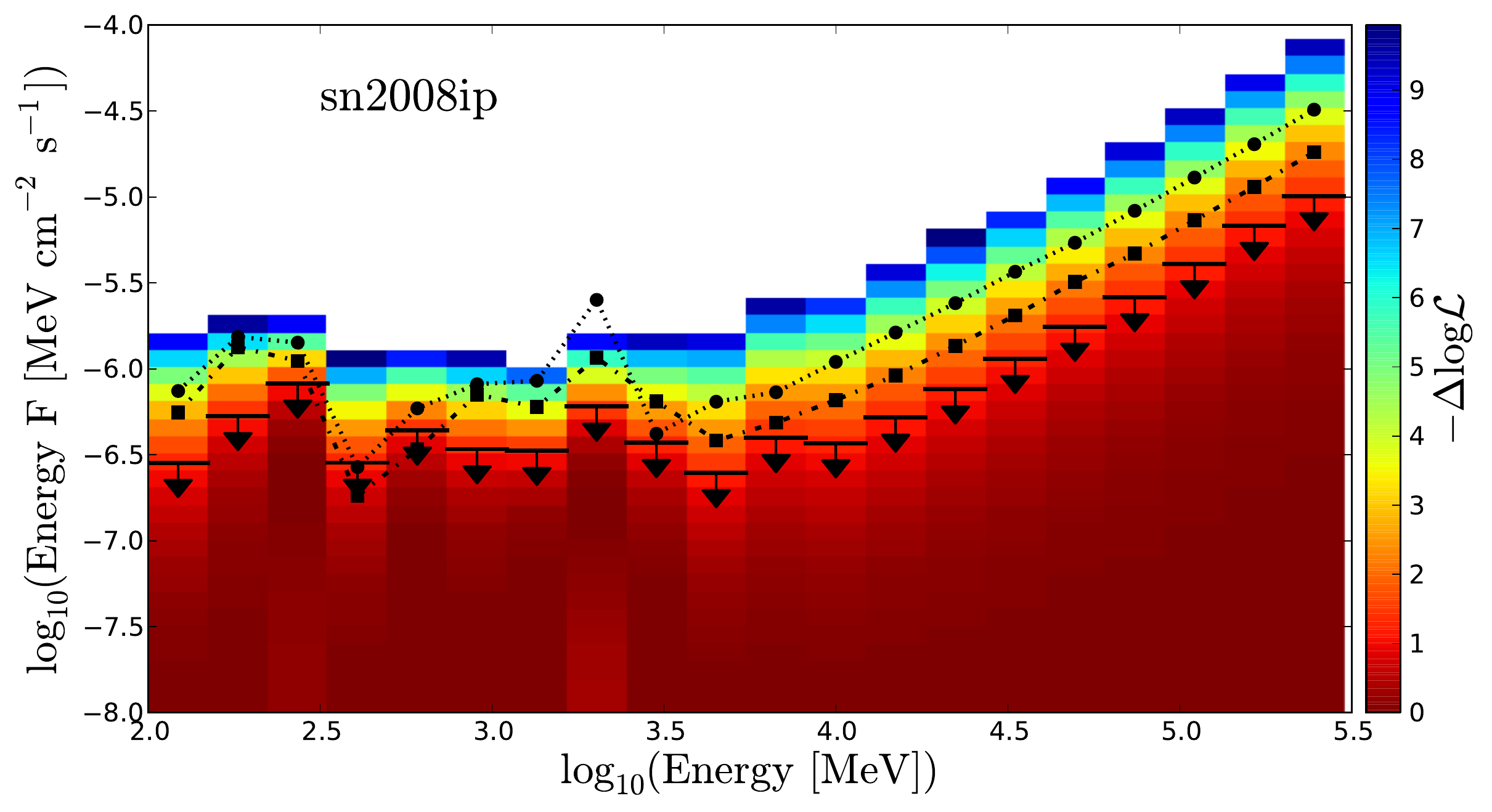}
\includegraphics[scale=\twopic]{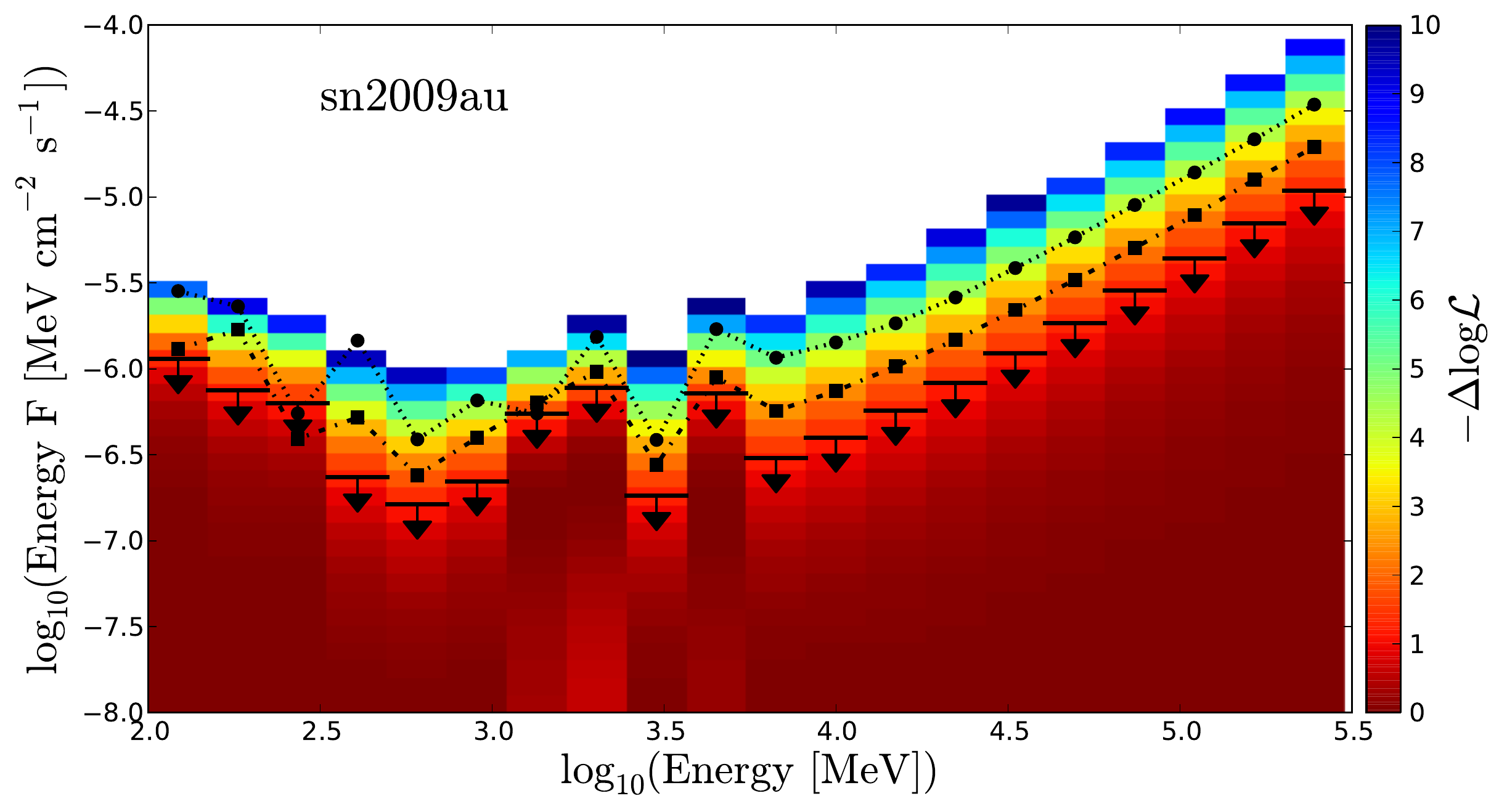}
\includegraphics[scale=\twopic]{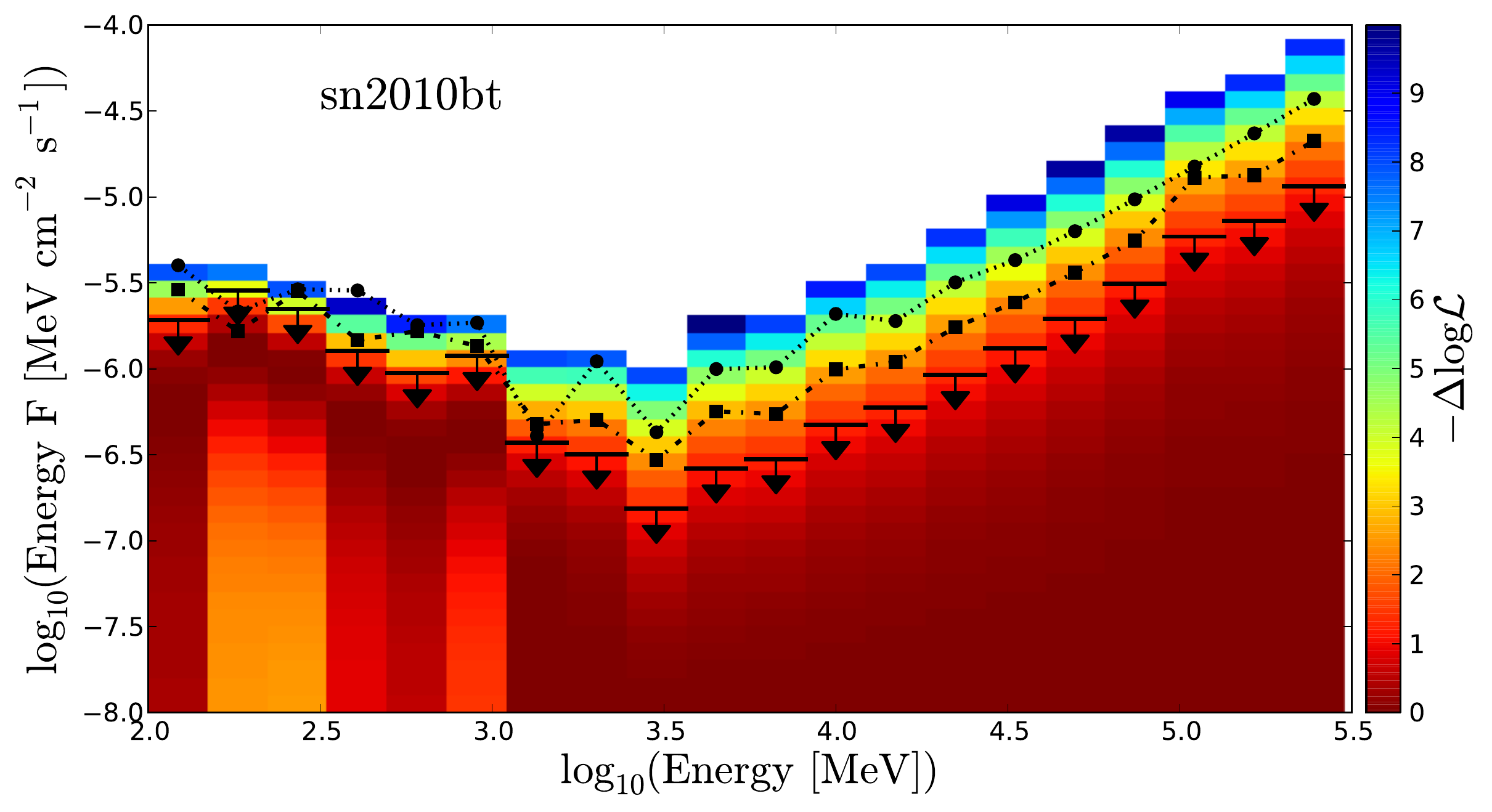}
\includegraphics[scale=\twopic]{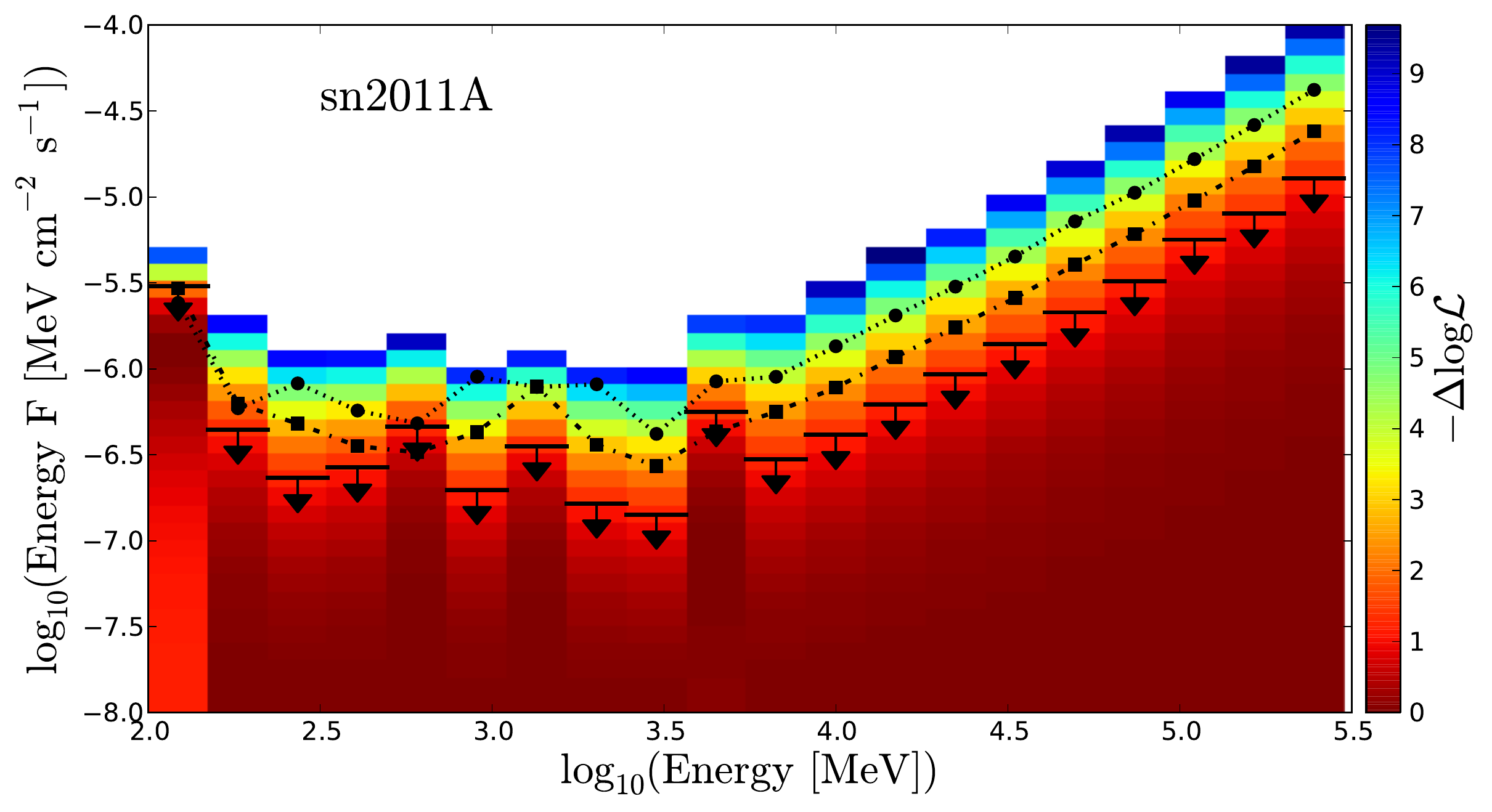}
\includegraphics[scale=\twopic]{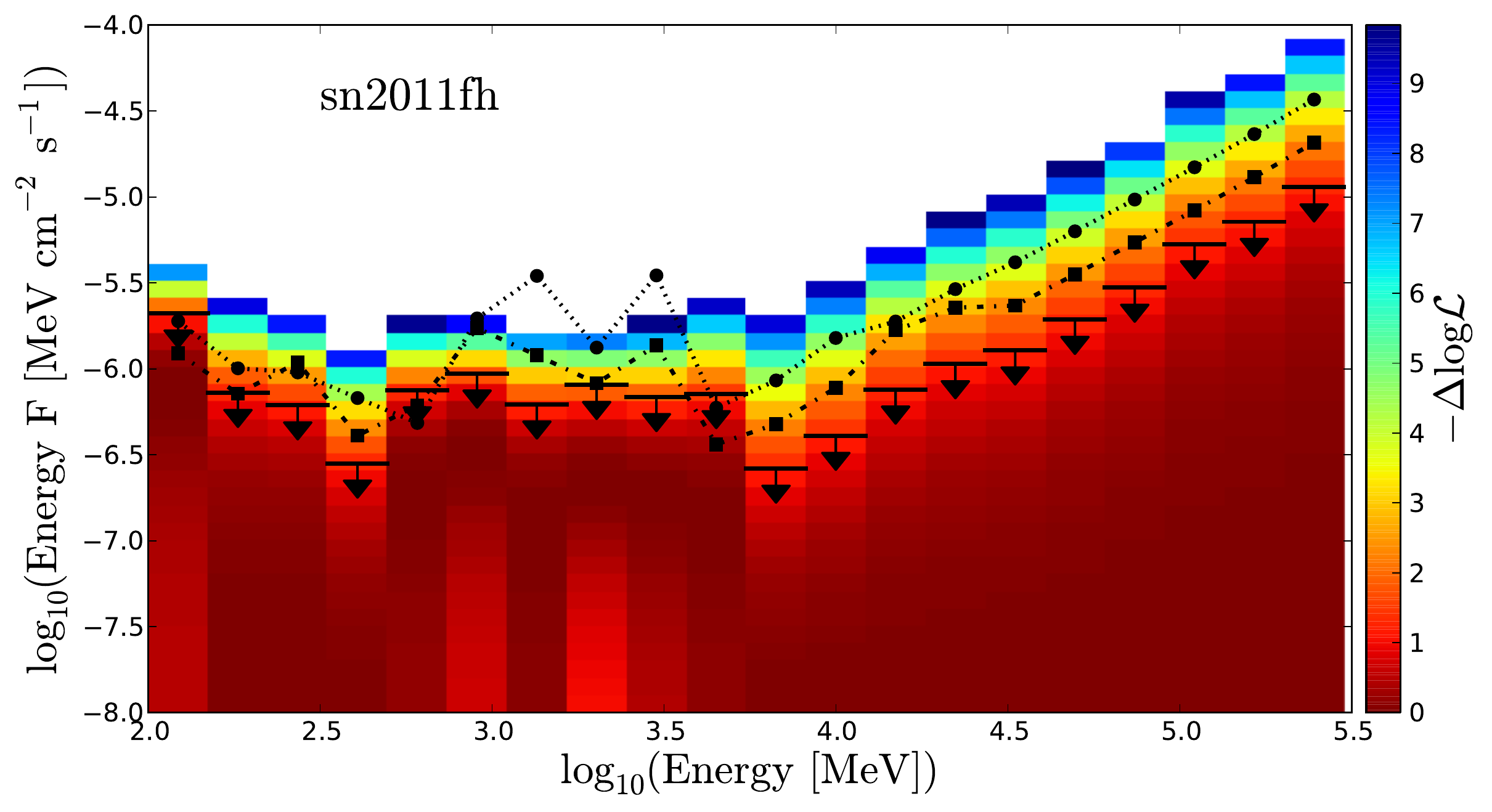}
\includegraphics[scale=\twopic]{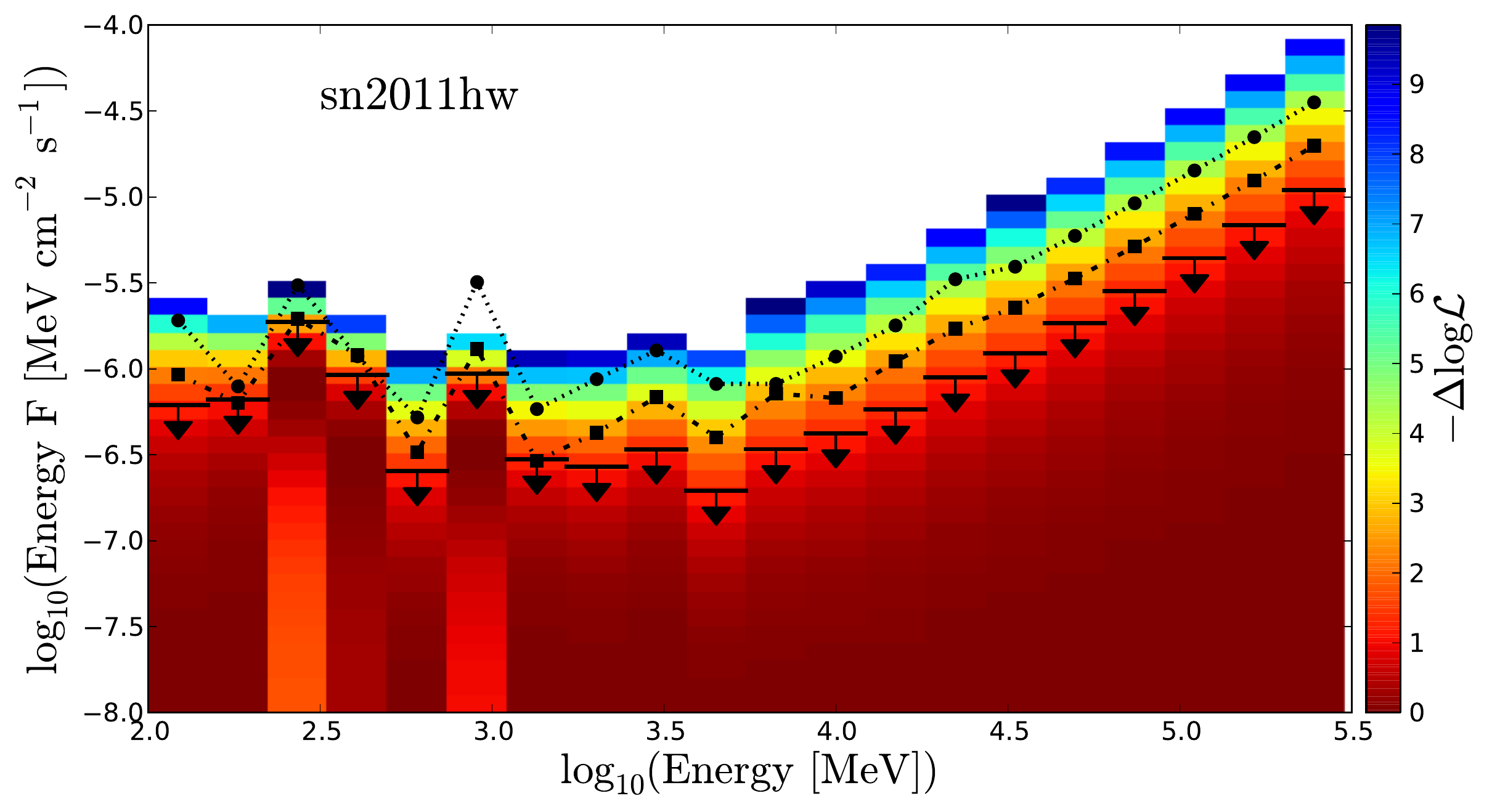}
\includegraphics[scale=\twopic]{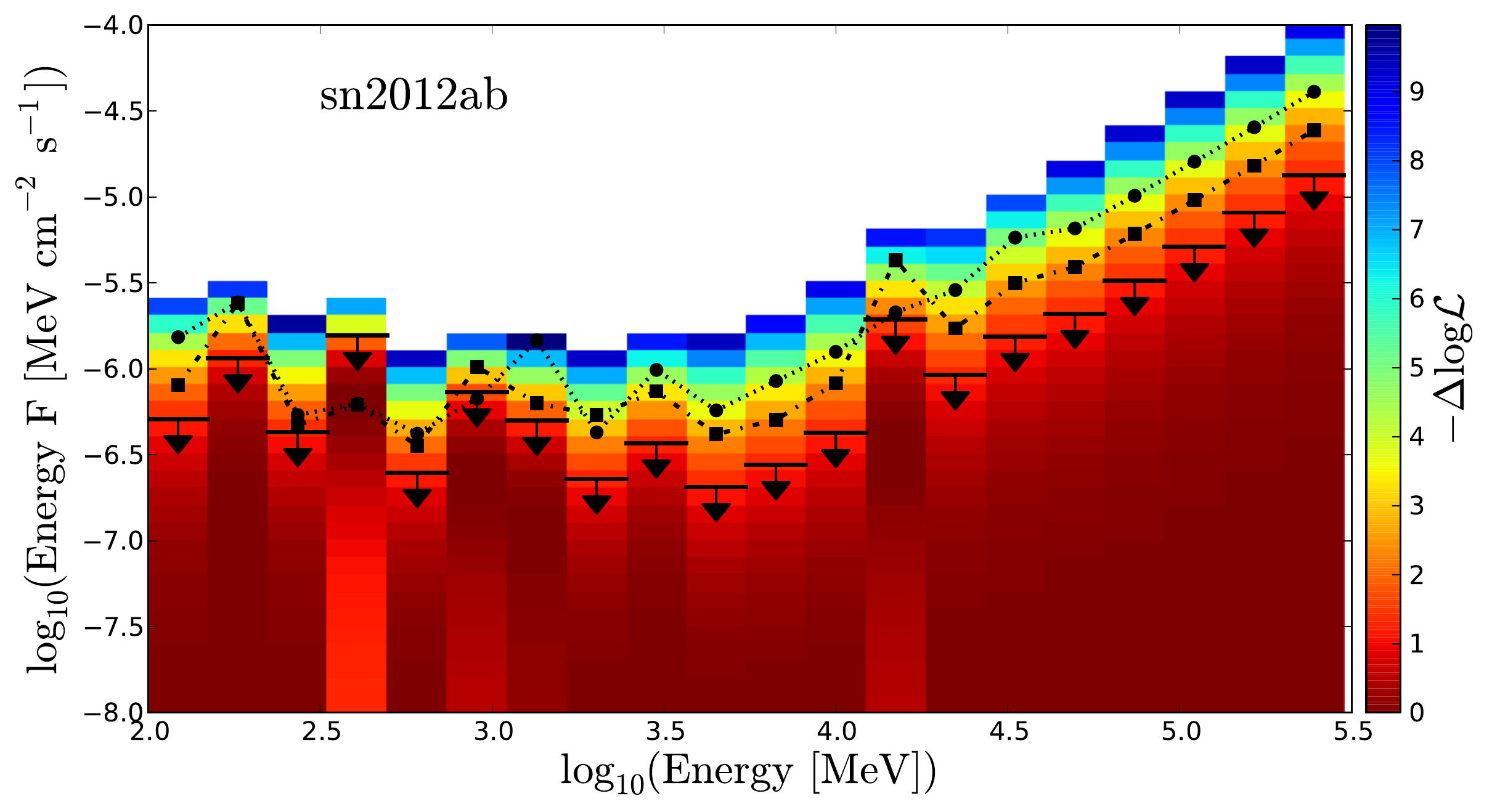}
\noindent
\caption{\small 
Similar to Fig.~\ref{fig:BinLLH}. Colors represent the likelihood profile for each energy for $\Delta T = 1$\,yr. The black arrows indicate the 95\% CL upper limits for $\Delta T = 1$\,yr, while the dotted-dashed and dotted lines represent the 95\% CL upper limit for $~\Delta T = 6$\,months and $\Delta T = 3$\,months respectively.}
\label{fig:BinLLHB}
\end{center}
\vspace{1mm}
\end{figure}

Figures~\ref{fig:BinLLHA} and~\ref{fig:BinLLHB} show the likelihood profiles in energy bins for $\Delta T = 1$\,yr and the 95\% CL upper limit for three time windows $\Delta T = 1$\,yr, $~\Delta T = 6$\,months and $\Delta T = 3$\,months for all SNe listed in Table~\ref{tab:closeSNshort}.

\newpage

\newpage
\bibliography{SN_paper}         


\end{document}